\documentclass[useAMS,usenatbib,usegraphicx]{mn2e}
\usepackage{bm}
\usepackage{times} 
\def\H2{H$_2$}
\def\sdss{SDSS J1048+4637}

\begin{document}
\title[Extinction curves of young galaxies]{Extinction curves
expected in young galaxies}
\author[H. Hirashita et al.]{Hiroyuki Hirashita$^{1,2}$\thanks{E-mail:
     hirashita@u.phys.nagoya-u.ac.jp}\thanks{Postdoctoral
     Fellow of the Japan Society for the Promotion of Science (JSPS).}, 
Takaya Nozawa$^{3}$,
Takashi Kozasa$^{3}$,
\newauthor
Takako T. Ishii$^{4}$\dag, and
Tsutomu T. Takeuchi$^{5}$\thanks{Postdoctoral Fellow of the JSPS
for Research Abroad.}
\newauthor
\\
$^1$ Graduate School of Science, Nagoya University, Nagoya
     464-8602, Japan \\
$^2$ SISSA/International School for Advanced Studies, Via
     Beirut 4, 34014 Trieste, Italy \\
$^3$ Division of Earth and Planetary Sciences, Graduate
     School of Science, Hokkaido University, Sapporo
     060-0810, Japan \\
$^4$ Kwasan Observatory, Kyoto University, Yamashina-ku,
     Kyoto 607-8471, Japan \\
$^5$ Laboratoire d'Astrophysique de Marseille, Traverse du Siphon BP 8, 13376
     Marseille Cedex 12, France
}
\date{Accepted 2004 December 10; Submitted 2004 September 8}
\pubyear{2005} \volume{000} \pagerange{1}
\twocolumn
%%\onecolumn

\maketitle \label{firstpage}
\begin{abstract}
We investigate the extinction curves of young galaxies in which
dust is supplied from Type II supernovae (SNe II) and/or pair
instability supernovae (PISNe). We adopt Nozawa et al.\ (2003)
for compositions and size distribution of grains formed in
SNe II
and PISNe. We find that the extinction curve is quite sensitive
to internal metal mixing in supernovae (SNe). The extinction
curves predicted from the mixed SNe are dominated by SiO$_2$
and is characterised by steep rise from infrared
to ultraviolet (UV). The dust from unmixed SNe shows
shallower extinction curve, because of the contribution from
large-sized ($\sim 0.1~\mu$m) Si grains. However, the
progenitor mass is important in unmixed SNe II: If the
progenitor mass is smaller than $\sim 20~M_\odot$,
the extinction curve is flat in UV; otherwise,
the extinction curve rises toward the short wavelength.
The extinction curve observed in a high-redshift quasar
($z=6.2$) favours the dust production by unmixed SNe II.
We also
provide some useful observational quantities, so that our
model might be compared with future high-$z$ extinction
curves.
\end{abstract}

\section{Introduction}

Dust grains play a crucial role in the formation and
evolution of galaxies. Dust grains absorb stellar light
and reemit it in far infrared (FIR), controlling the
energy balance in the interstellar medium (ISM) and
protostellar gas clouds. Also, the surface of dust grains
is a site for an efficient formation of \H2 molecules
\citep[e.g.][]{cazaux02,cazaux04}, which act as an
effective coolant in metal-poor ISM. Those effects of
dust turn on even in a very metal-poor environment
($\sim 1$\% of the solar metallicity;
Hirashita \& Ferrara 2002; Morgan \& Edmunds 2003), and
it is argued that the star formation rate is enhanced because
of the first dust enrichment in the history of galaxy
evolution. The first source of dust in the universe is Type II
supernovae (SNe II) or pair instability supernovae
(PISNe), since the lifetime of their progenitors is short
($\sim 10^6$ yr). In the local universe, dust grains are
also produced by evolved low mass stars (Gehrz 1989), but
this production mechanism requires much longer
($\ga 1$ Gyr) timescales. The first dust supplied by
SNe II or PISNe may trigger the
formation of low-mass stars (Schneider et al.\ 2003).

Since dust is important even in the early stage of
galaxy evolution, it is crucial to know how much dust
and what species of dust form. Todini \& Ferrara (2001),
following the method by
Kozasa, Hasegawa, \& Nomoto (1989, 1991), show that dust
mass produced by a SN II is roughly 0.1--0.4 $M_\odot$.
They also find that SNe II form amorphous carbon with
size around 300 \AA\ and silicate grains around
10--20 \AA. Schneider, Ferrara, \& Salvaterra (2004)
extend the progenitor mass range to the regime of PISNe
(140--260 $M_\odot$) and find that 30--60 $M_\odot$ of
dust forms per PISN. The grain radii are distributed
from 0.001 to 0.3 $\mu$m, depending on the species.
The motivation for considering PISNe comes from some
evidence indicating that the stars formed from
metal-free gas, Population III (PopIII) stars, are very
massive with a characteristic mass of a few hundred
solar masses (Bromm \& Larson 2004, and references
therein). Such massive stars are
considered to begin pair creation of electron and
positron after the helium burning phase, and finally
end their lives with explosive nuclear reaction
disrupting the whole stars
(Fryer, Woosley, \& Heger 2001). This explosion is
called PISN.

Nozawa et al.\ (2003) also calculate the dust mass in
the ejecta of PopIII SNe II and PISNe, carefully
treating the radial density profile and the
temperature evolution. Since it is still debated how
efficiently the mixing of atoms within supernovae (SNe)
occurs, Nozawa et al.\ (2003) treat two extreme cases
for the mixing of elements: one is the {\it unmixed}
case in which the original onion-like structure of
elements is preserved, and the other is the
{\it mixed} case in which the elements are uniformly
mixed within the helium core. After examining those
two cases, they show that the formed dust species
depend largely on the mixing of seed elements within
SNe, because the dominant reactions change depending on
the ratio of available elements (see
Section \ref{subsec:nozawa} for more detailed
description). Nozawa et al.\ (2003)
predict larger dust mass than Todini \& Ferrara (2001)
for SNe II.
How much and what kind of grain species form in SNe II
and PISNe is still a matter of debate, partly
depending on the degree of mixing within the He-core,
and on the model of SNe employed in the
calculations; Woosley \& Weaver (1995) for SNe II
in Todini \& Ferrara (2001), Heger \& Woosley (2002)
for PISNe in Schneider et al.\ (2004), and
Umeda \& Nomoto (2002) for PopIII
SNe in Nozawa et al.\ (2003).
The formed grain species in the calculation of
Nozawa et al.\ (2003) are listed in
Table \ref{tab:species}.

\begin{table}
\centering
\begin{minipage}{80mm}
\caption{Summary of grain species.}
\begin{tabular}{@{}lccc@{}}\hline
Species & condition\footnote{The classifications ``m'', ``u'',
and ``m/u'' mean that the
species is formed in mixed, unmixed, and both supernovae,
respectively.} & Ref\footnote{References for optical
constants: (1) Edo (1983); (2) Edward (1985);
(3) Philipp (1985); (4) Lynch \& Hunter (1991);
(5) Semenov et al.\ (2003); (6) Mukai (1989) (for
$\lambda <0.14~\mu$m, we adopt the values at
$\lambda =0.14~\mu$m);
(7) Toon, Pollack, \& Khare (1976);
(8) Roessler \& Huffman (1991);
(9) Dorschner et al.\ (1995);
(10) J\"{a}ger et al.\ (2003).}
&  density ($\delta_j$) \\
& & & (g cm$^{-3}$)
%%\multicolumn{2}{c}{[$M_\odot~{\rm yr}^{-1}~{\rm kpc}^{-2}$]}
 \\ \hline
C  & u & 1 & 2.28 \\
Si & u & 2 & 2.34 \\
SiO$_2$& m/u & 3 & 2.66 \\
Fe & u & 4 & 7.95 \\
FeS & u & 5 & 4.87 \\
Fe$_3$O$_4$ & m & 6 & 5.25 \\
Al$_2$O$_3$ & m/u & 7 & 4.01 \\
MgO     & u & 8 & 3.59 \\
MgSiO$_3$ & m/u & 9 & 3.20 \\
Mg$_2$SiO$_4$ & m/u & 10 & 3.23 \\
\hline
\end{tabular}
\label{tab:species}
\end{minipage}
\end{table}

Some observations have detected infrared
radiation from extragalactic SNe II (e.g.\
Dwek et al.\ 1983;
Moseley et al.\ 1989;
Kozasa, Hasegawa, \& Nomoto 1989). This radiation
has been interpreted to be originating from dust
formed in SNe II. FIR and sub-millimetre (sub-mm)
observations of Galactic SN remnants also have
recently put further constraints on the dust mass
formed in SNe II
(Cas A: e.g.\ Arendt, Dwek, \& Moseley 1999;
Dunne et al.\ 2003; Hines et al.\ 2004;
Kepler: e.g.\ Morgan et al.\ 2003). Although
FIR and sub-mm observations are useful to know
the dust amount, the emissivity, which reflects the
composition, and the dust amount are degenerated in
the observed FIR and sub-mm luminosity.
Therefore, in order to constrain the model of dust
formation in SNe II, another independent
information on the dust amount and composition is
necessary.

Extinction curves are often used to investigate
the dust properties (e.g.\ Mathis 1990).
Recently, by using a sample of broad absorption
line (BAL) quasars, Maiolino et al.\ (2004a) have
shown that the extinction properties of the
low-redshift ($z<4$, where $z$ is the redshift)
sample is different from those of the high-$z$
($z>4.9$) sample. This result is suggestive of a
change in the dust production mechanism in the
course of galaxy evolution. The highest-redshift
BAL quasar in their sample,
SDSS J104845.05+463718.3 (hereafter \sdss) at
$z=6.2$ shows a red spectrum at the restframe
wavelength $\lambda <1700$ \AA. (In this paper,
all the
wavelengths are shown in the restframe of observed
galaxies.)
However, at $\lambda >1700$ \AA, there is no
indication of reddening. Then they suggest
the extinction curve to be flat at
$\lambda \ga 1700$ \AA\ and rising at
$\lambda\la 1700$ \AA.
Maiolino et al.\ (2004b) find that
the extinction curve
of \sdss\ is different from that of low-$z$ BAL
quasars and is
in excellent agreement with the SN II dust models by
Todini \& Ferrara (2001). It is interesting to extend
their work to various dust formation models in
Nozawa et al.\ (2003). 
The extinction curve should
be different from SNe II to PISNe and from mixed SNe to
unmixed SNe.
Dust production in such various
conditions is extensively investigated in
Nozawa et al.\ (2003);
hence we aim at investigating the extinction
curve based on their results.

In addition to Maiolino et al.\ (2004a), evidence for
dust enrichment has been obtained at very high $z$
($>5$), where the cosmic age is less than 1 Gyr,
by the recent sub-mm and millimetre observations of
distant quasars (Bertoldi et al.\ 2003;
Priddey et al.\ 2003). At lower $z$ ($\la 5$),
direct indications of high-$z$ dust comes from the
reddening of background quasars
(Fall, Pei, \& McMahon 1989; Zuo et al.\ 1997). 
The depletion of heavy elements in quasar absorption line
systems also supports the presence of dust in distant
systems (Pettini et al.\ 1994;
Molaro, Vladilo, \& Centurion 1998;
Levshakov et al.\ 2000; Vladilo 2002;
Ledoux, Petitjean, \& Srianand 2003).
There are several observations of extinction curves
up to $z\sim 1$ by taking
advantage of the gravitational lensing
(Falco et al.\ 1999; and Mu\~{n}oz et al.\ 2004).
Spectropolarimetric observations of two radio galaxies at
$z\sim 1.4$ reveal the 2200-\AA\ dust feature in scattered
light (Sol\'{o}rzano-I\~{n}arrea et al.\ 2004).
However, most of the observations of extinction curves are
limited to
relatively low $z$, where dust is not only produced by
SNe II, but  also by evolved late-type stars.
In the future, observational samples of extinction curves
could be extended to high-$z$ primeval galaxies,
where dust is predominantly produced in SNe II and/or
PISNe. This work could be applied to
such future observations to reveal the size and
composition of dust originating from SNe II or PISNe.

We first describe our theoretical treatment to calculate the
extinction curves of SN II and PISN dust in
Section \ref{sec:model}. We examine our results, and
provide some observationally convenient quantities in
Section \ref{sec:results}. We discuss our results from the
observational viewpoint in Section \ref{sec:obs}, and
finally give the conclusion of this paper in
Section \ref{sec:sum}.

\section{Model}\label{sec:model}

Our aim is to derive the theoretical extinction curves of
dust produced in SNe II and PISNe. We adopt the dust
production model by Nozawa et al.\ (2003), who investigate
various progenitor mass of SNe II and PISNe with a careful
treatment of physical processes (internal mixing,
temperature evolution, etc.).

\subsection{Dust production in SNe II and PISNe}
\label{subsec:nozawa}

Nozawa et al.\ (2003) investigate the formation of dust
grains in the ejecta of PopIII SNe (SNe II and
PISNe, whose progenitors are initially metal-free). The
calculation treats some details compared with
Todini \& Ferrara (2001): (i) the radiative transfer
equation including the energy deposition of radioactive
elements is solved to calculate the time evolution of
gas temperature, which strongly affects the number density
and size of newly formed grains; (ii) the radial profile
of density of various metals is considered;
(iii) unmixed and uniformly mixed cases
within the helium core are considered.

In the unmixed case, Nozawa et al.\ (2003) assume that
the original onion-like structure of elements is
preserved. On the other hand, in the mixed case,
they uniformly mix all the elements in the helium
core. They also assume the complete formation of
CO and SiO molecules, neglecting the destruction of
those molecules: no carbon-bearing grain condenses in
the region of
${\rm C/O}<1$ and no Si-bearing grain, except for
oxide grains, condenses in the region of
${\rm Si/O}<1$.
The formation of CO and SiO
may be incomplete because of the destruction by
energetic electron impact within
SNe. Todini \& Ferrara (2001) treat both formation
and destruction of CO and SiO, finding that
both are mostly destroyed. The decrease
of CO leads to the formation of
carbon grains, which could finally be
oxidised with available oxygen. The destruction of
SiO could decrease the formation of grains
composed of SiO$_2$, MgSiO$_3$, and Mg$_2$SiO$_4$,
and increase other oxidised grains and Si grains.
Observationally, it is still debated
if CO and SiO are efficiently destroyed or not.
A detailed discussion on this issue can be found in
Appendix B of Nozawa et al.\ (2003).

In the unmixed ejecta, a variety of grain species
(Si, Fe, Mg$_2$SiO$_4$, MgSiO$_3$, MgO, Al$_2$O$_3$,
SiO$_2$, FeS, and C) condense, while in the mixed ejecta,
only oxide grains (SiO$_2$, MgSiO$_3$, Mg$_2$SiO$_4$,
Al$_2$O$_3$, and Fe$_3$O$_4$) form. The species are
summarised in Table \ref{tab:species}. The difference
in the formed species between mixed and unmixed cases
are mainly derived by the
formation of molecules. The
carbon dust is not produced in
their mixed case, because the carbon and
oxygen are mixed and combined
to form CO molecules. On the contrary,
it forms in
unmixed SNe, since there is a carbon-rich region
at a certain radius of SNe. The formation of SiO
molecules also affects the formed species:
in the mixed ejecta only oxide silicate grains
form, while in the unmixed ejecta non-oxide
grains can form in oxygen-poor regions.

The size of the grains on the location of formation
site in the ejecta spans a range of 3 orders of
magnitude, depending on the grain
species. The size distribution function summed up over
all the grain species is approximated by a broken
power law. This size distribution is different from that
of the SN II calculation of Todini \& Ferrara (2001),
which has
a typical sizes of 300 \AA\ for amorphous carbon and
10--20 \AA\ for oxide grains. The difference is mainly
comes from the different treatment of the ejecta:
Nozawa et al.\ (2003) consider the density and
temperature structures within the helium core, while
Todini \& Ferrara (2001) do not.
Schneider et al.\ (2004), based on
the model of Todini \& Ferrara (2001), find for
PISNe a large range of dust size, depending on the
species. Their result is similar to the mixed
case in Nozawa et al.\ (2003).

We adopt the representative progenitor mass of SNe II
as 20 $M_\odot$ and that of PISNe as 170 $M_\odot$.
The size distribution of each grain species is almost
independent of the progenitor mass, if the SN type
is fixed (SN II or PISN), except for unmixed SNe II
(Section \ref{subsec:theor}). Therefore, we
concentrate only those two masses in this paper.
However, the relative mass ratio among species
mildly depends on
the progenitor mass, and we comment on it later
showing the contribution of each species to the
extinction curve (Section \ref{subsec:theor}). We
also investigate the mixed and unmixed cases.
Therefore, we treat four cases: (a) mixed SNe II;
(b) unmixed SNe II; (c) mixed PISNe; (d) unmixed
PISNe, as summarised in Table \ref{tab:cases}.
All the formulation and the results can be seen in
Nozawa et al.\ (2003).
In this paper, we assume the grains to be uniform
and spherical.

\begin{table*}
\centering
\begin{minipage}{160mm}
\caption{Models of dust production in supernovae.}
\begin{tabular}{@{}cccccc@{}}\hline
Model & Progenitor mass & Mixing & $R_V$ &
$\langle\sigma_{\rm d}(V)/m_{\rm d}\rangle$ &
$\langle\sigma_{\rm d}(0.3~\mu{\rm m})/m_{\rm d}\rangle$ \\
 & ($M_\odot$) & & & \multicolumn{2}{c}{($10^4$ cm$^2$ g$^{-1}$)} \\
\hline
a &  20 & mixed   & 2.4 & 0.98 & 2.8 \\
b &  20 & unmixed & 3.3 & 2.2  & 4.4 \\
c & 170 & mixed   & 1.4 & 0.75 & 3.0 \\
d & 170 & unmixed & 5.0 & 2.1  & 4.1  \\
Galactic\footnote{Quantities derived from observational properties
of the Galactic extinction properties (Spitzer 1978;
Cardelli et al.\ 1989; Mathis 1990). We should note that it is
not necessary to explain the Galactic properties with our models,
because the origin, composition, and size of dust are different.}
& --- & --- & 3.1 & 3.4 & 6.3 \\
\hline
\end{tabular}
\label{tab:cases}
\end{minipage}
\end{table*}

\begin{figure*}
\includegraphics[width=8cm]{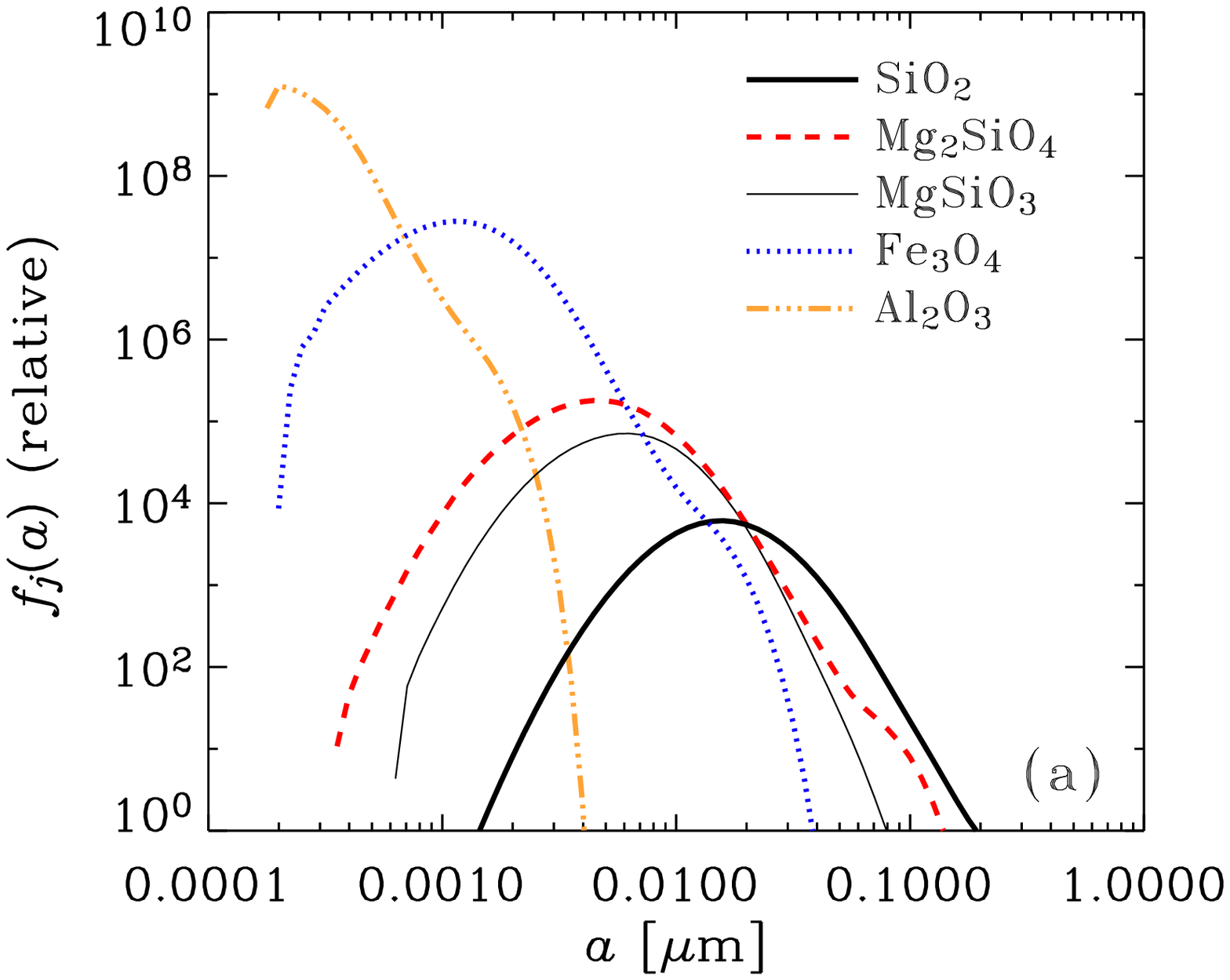}
\includegraphics[width=8cm]{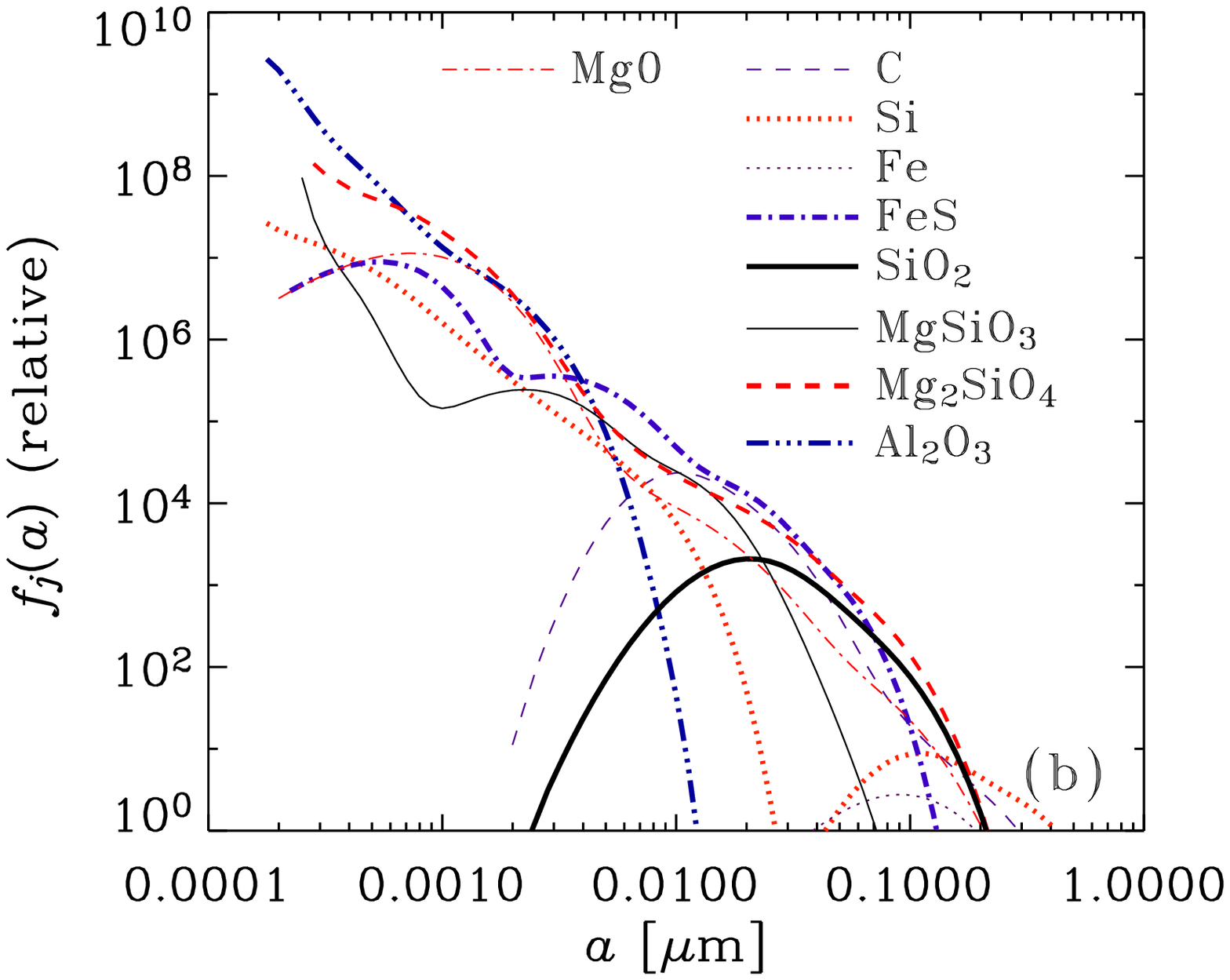}
\includegraphics[width=8cm]{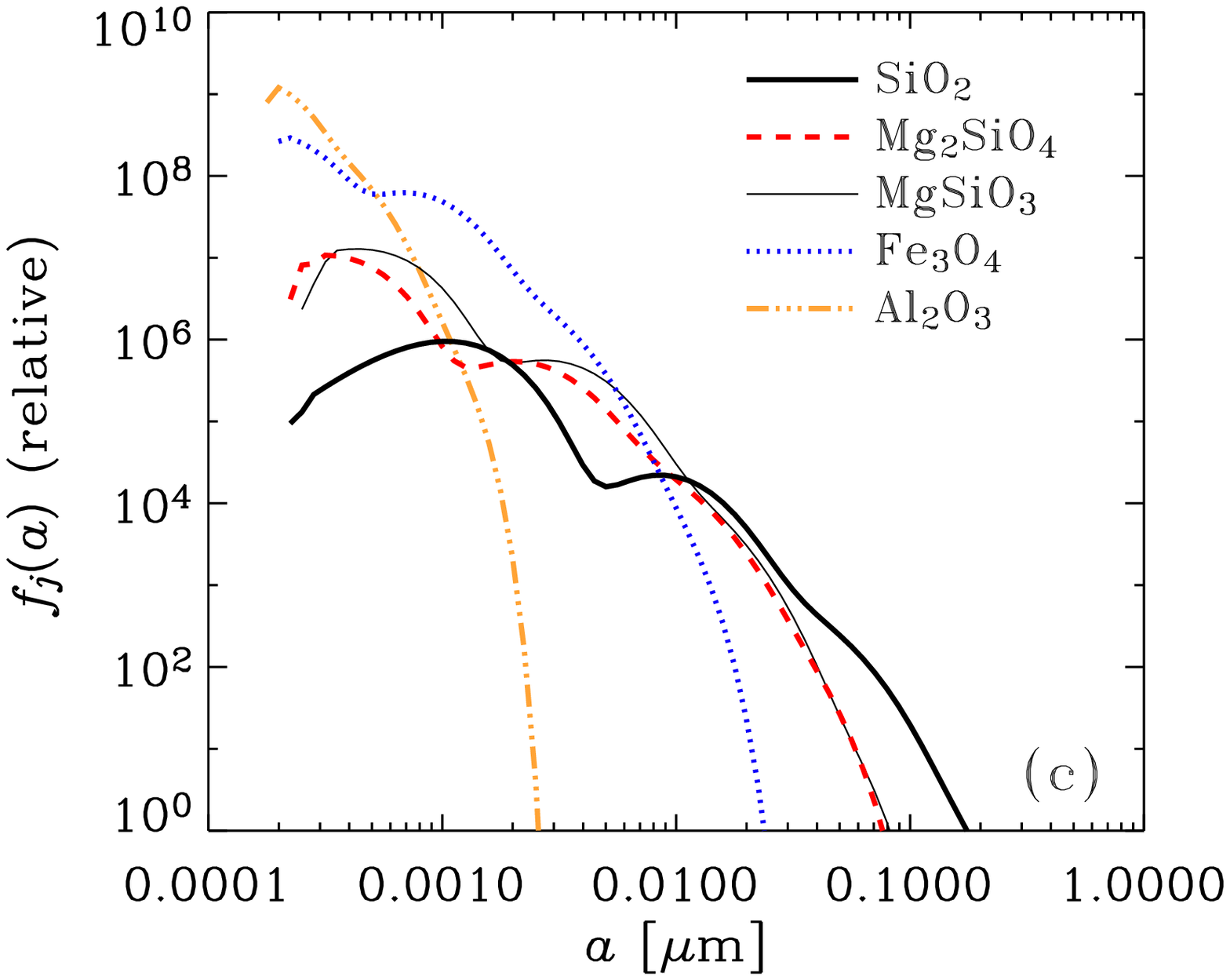}
\includegraphics[width=8cm]{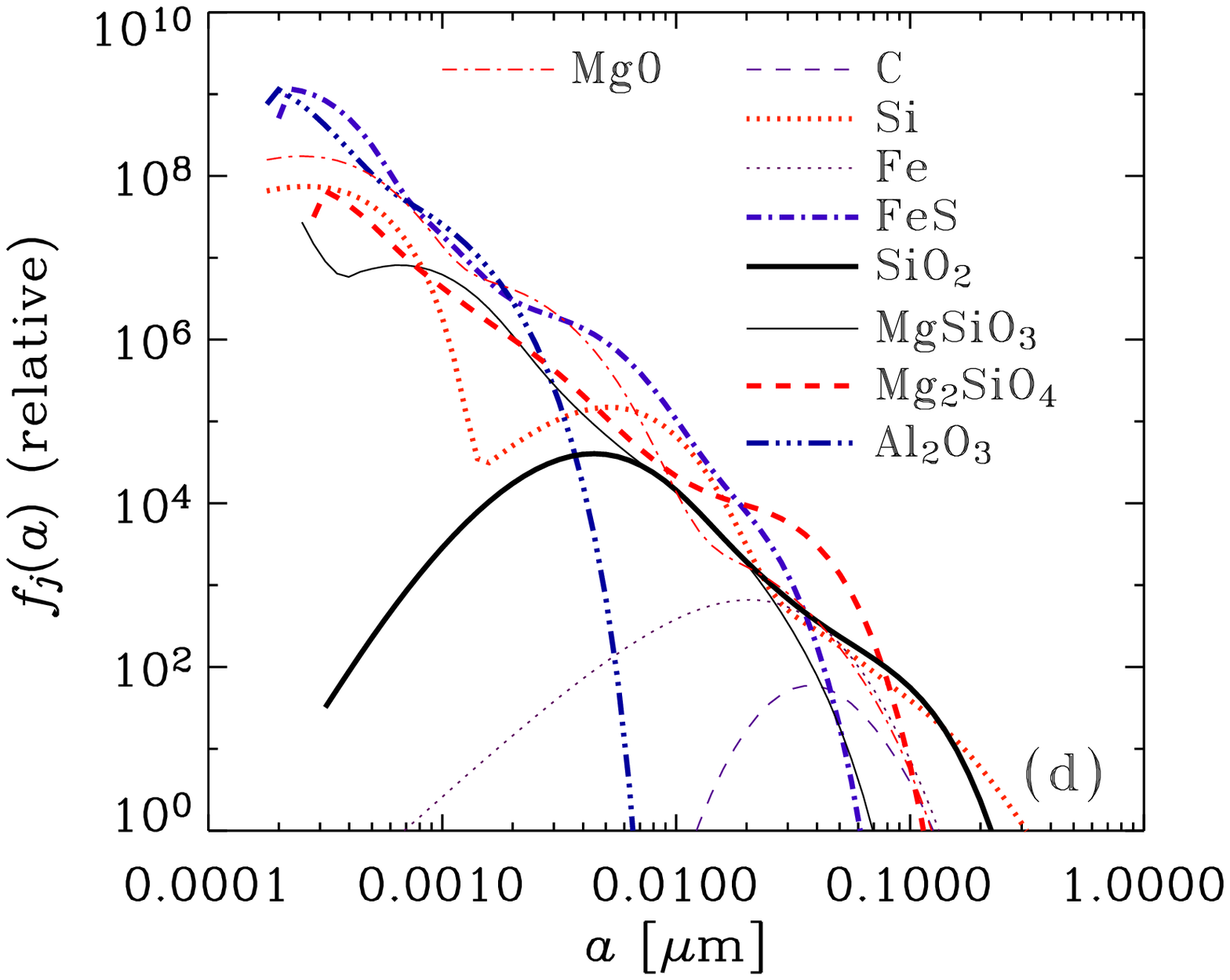}
\caption{Size distribution function of each grain species
in (a) the mixed ejecta with the progenitor of 20 $M_\odot$,
(b) the unmixed ejecta with the progenitor of 20 $M_\odot$,
(c) the mixed ejecta with the progenitor of 170 $M_\odot$,
and (d) the unmixed ejecta with the progenitor of
170 $M_\odot$. The correspondence between the species and
lines is shown in the figures.
\label{fig:size}}
\end{figure*}

In Figure \ref{fig:size}, we show the size
distribution adopted in this paper, where the size
distribution function $f_j(a)$ is defined so that
$f_j(a){\rm d}a$ is proportional to the number of
grains in a radius interval $[a,\, a+{\rm d}a]$
($j$ indicates the species). The four figures
correspond to the four cases in
Table \ref{tab:cases}.
The normalisation of $f_j(a)$ is discussed in
Section \ref{subsec:useful}, and we only apply an
arbitrary normalisation to each figure.

\subsection{Calculation of Extinction Curves}
\label{subsec:method}

In order to calculate extinction curves, the optical
constants for the grains are necessary. We adopt the
references listed in Table \ref{tab:species} for the
optical constants.
By using those optical constants, we calculate the
absorption and scattering cross sections of homogeneous
spherical grains with various sizes based on the Mie
theory (Bohren \& Huffman 1983). The efficiency factor
of extinction,
which is defined as the cross section divided by the
geometrical cross section, is denoted as
$Q_{{\rm ext},\, j}(\lambda,\, a)$. This efficiency
factor is a function of
the wavelength $\lambda$ and the dust size $a$
($j$ denotes the grain species).

The optical depth of grain $j$ as a function of
wavelength, $\tau_{\lambda ,\, j}$, is calculated by
weighting the cross sections according to the size
distribution as shown in Figure \ref{fig:size}:
\begin{eqnarray}
\tau_{\lambda ,\, j}=\int_0^\infty\pi a^2
Q_{{\rm ext},\, j}(\lambda ,\, a)\, Cf_j(a)\,
{\rm d}a\, ,\label{eq:tau}
\end{eqnarray}
where $C$ is the
normalisation constant related to the column density
of dust. The determination of $C$ is not important for
the extinction curve because the extinction 
curve is presented in the form of
$A_\lambda /A_V$ ($A_\lambda$ is the extinction in
units of magnitude at wavelength $\lambda$, and the
$V$-band wavelength is 0.55 $\mu$m), where the
constant $C$ is canceled.
The determination of $C$ is necessary when we quantify
the column density (equation \ref{eq:norm}). The
extinction in units of
magnitude is proportional to the optical depth as
\begin{eqnarray}
A_{\lambda,\, j}=1.086\tau_{\lambda ,\, j}
\, ,\label{eq:extj}
\end{eqnarray}
where $A_{\lambda,\,j}$ is the extinction of species
$j$ in units of magnitude as a function of $\lambda$,
and the factor 1.086 comes from $2.5\log_{10}e$. The
total extinction $A_\lambda$ is
calculated by summing $A_{\lambda,\,j}$ for all the
concerning species:
\begin{eqnarray}
A_\lambda =\sum_j A_{\lambda ,\, j}\, .
\label{eq:extall}
\end{eqnarray}

\section{RESULTS}\label{sec:results}

\subsection{Extinction curves of various SNe}
\label{subsec:theor}

\begin{figure*}
\includegraphics[width=8cm]{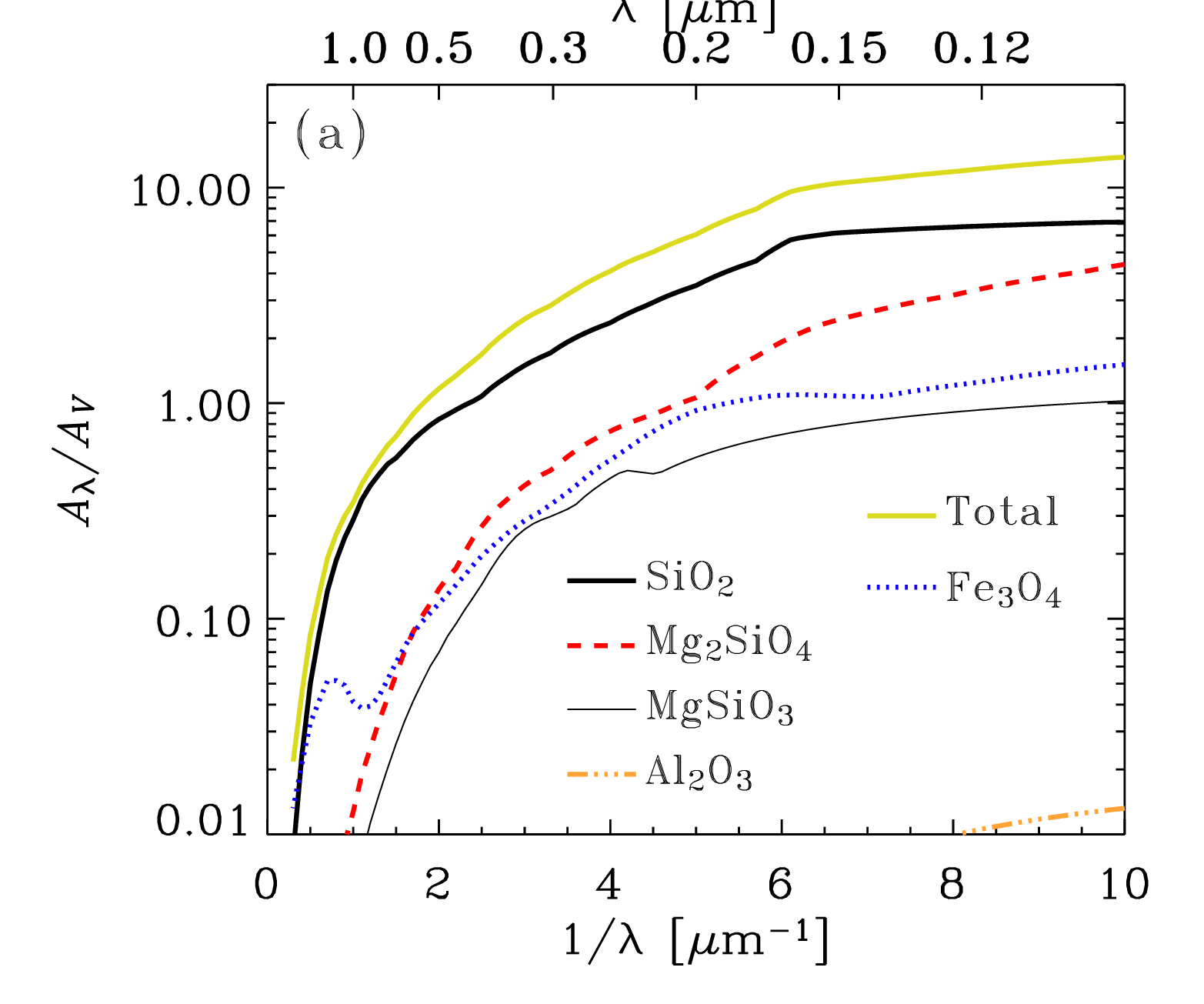}
\includegraphics[width=8cm]{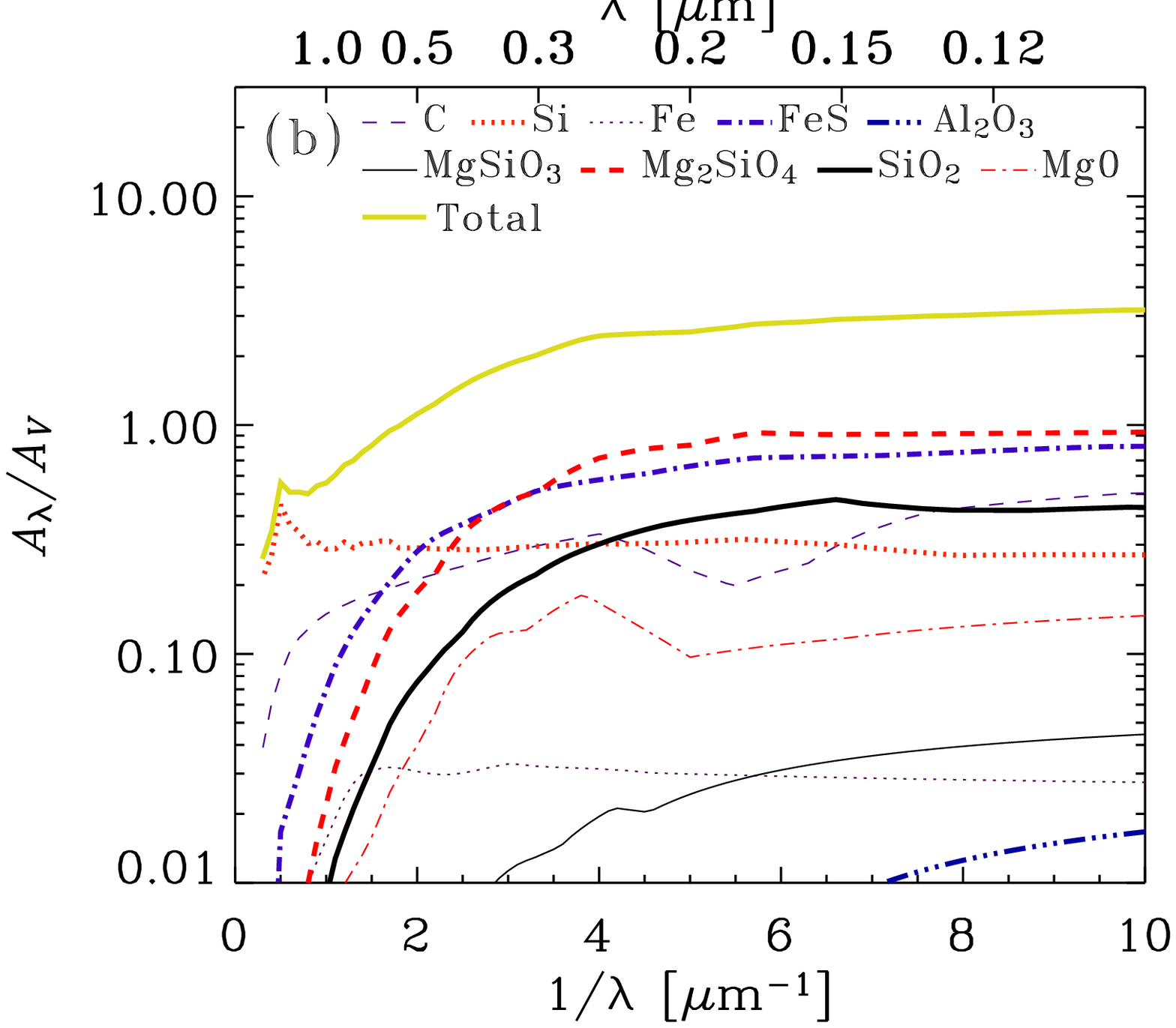}
\includegraphics[width=8cm]{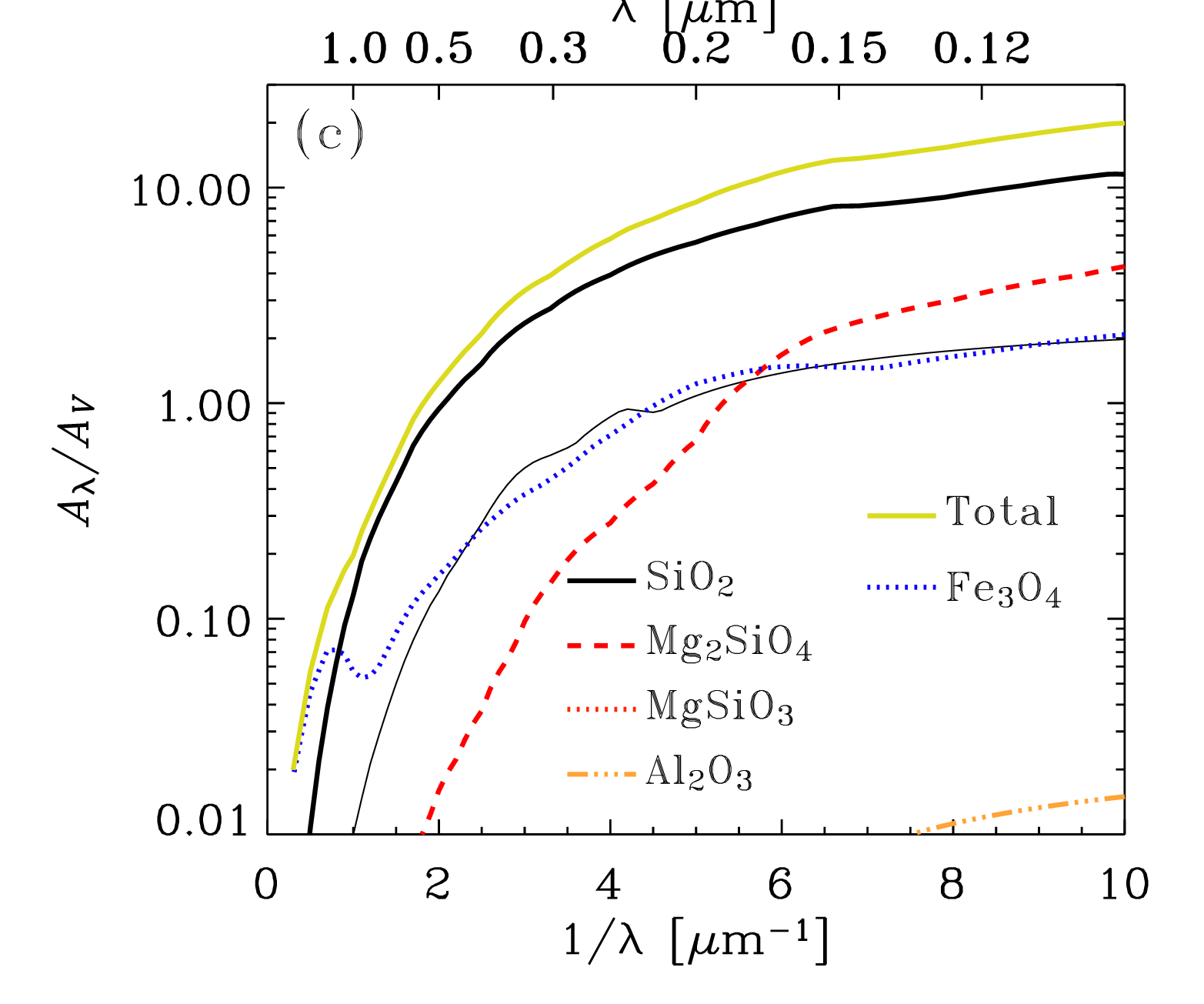}
\includegraphics[width=8cm]{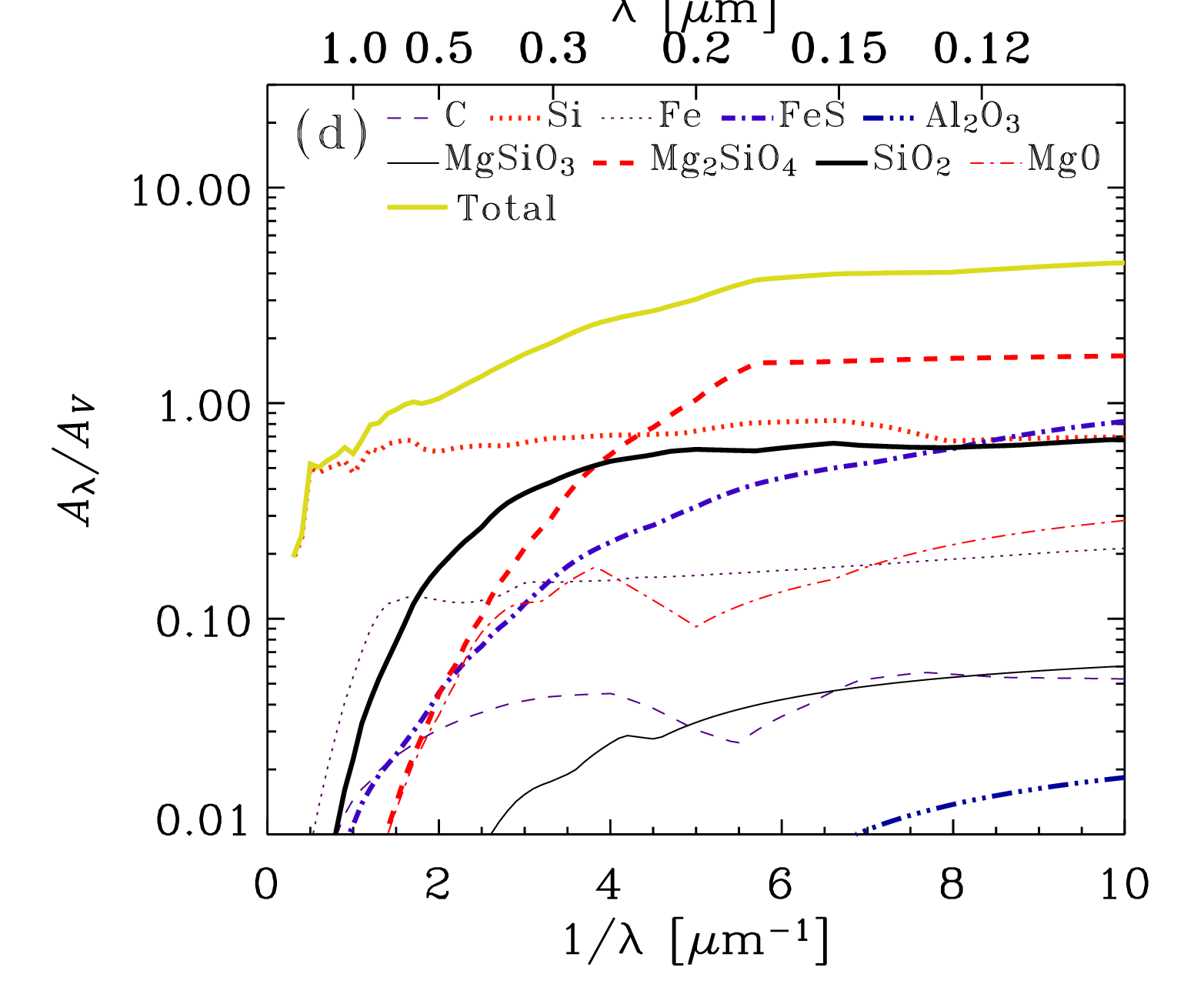}
\caption{Extinction curves calculated for the dust
production models listed in Table \ref{tab:cases}: (a)
the mixed ejecta with the progenitor of 20 $M_\odot$;
(b) the unmixed ejecta with the progenitor of
20 $M_\odot$; (c) the mixed ejecta with the
progenitor of 170 $M_\odot$; and (d) the unmixed
ejecta with the progenitor of 170 $M_\odot$. The
contribution of each species is also shown. Each curve
is normalised to the $V$-band ($\lambda =0.55~\mu$m)
value of the total extinction curve. The
correspondence between the species and lines is shown
in the figures.
\label{fig:theoretical}}
\end{figure*}

In Figure \ref{fig:theoretical}a, we show the extinction
curves of the four cases in Table \ref{tab:cases}:
(a) mixed SNe II; (b) unmixed SNe II; (c) mixed PISNe;
(d) unmixed PISNe. The contribution of each species is
also shown. The extinction curve is normalised to
$A_V$.

\begin{figure*}
\includegraphics[width=8cm]{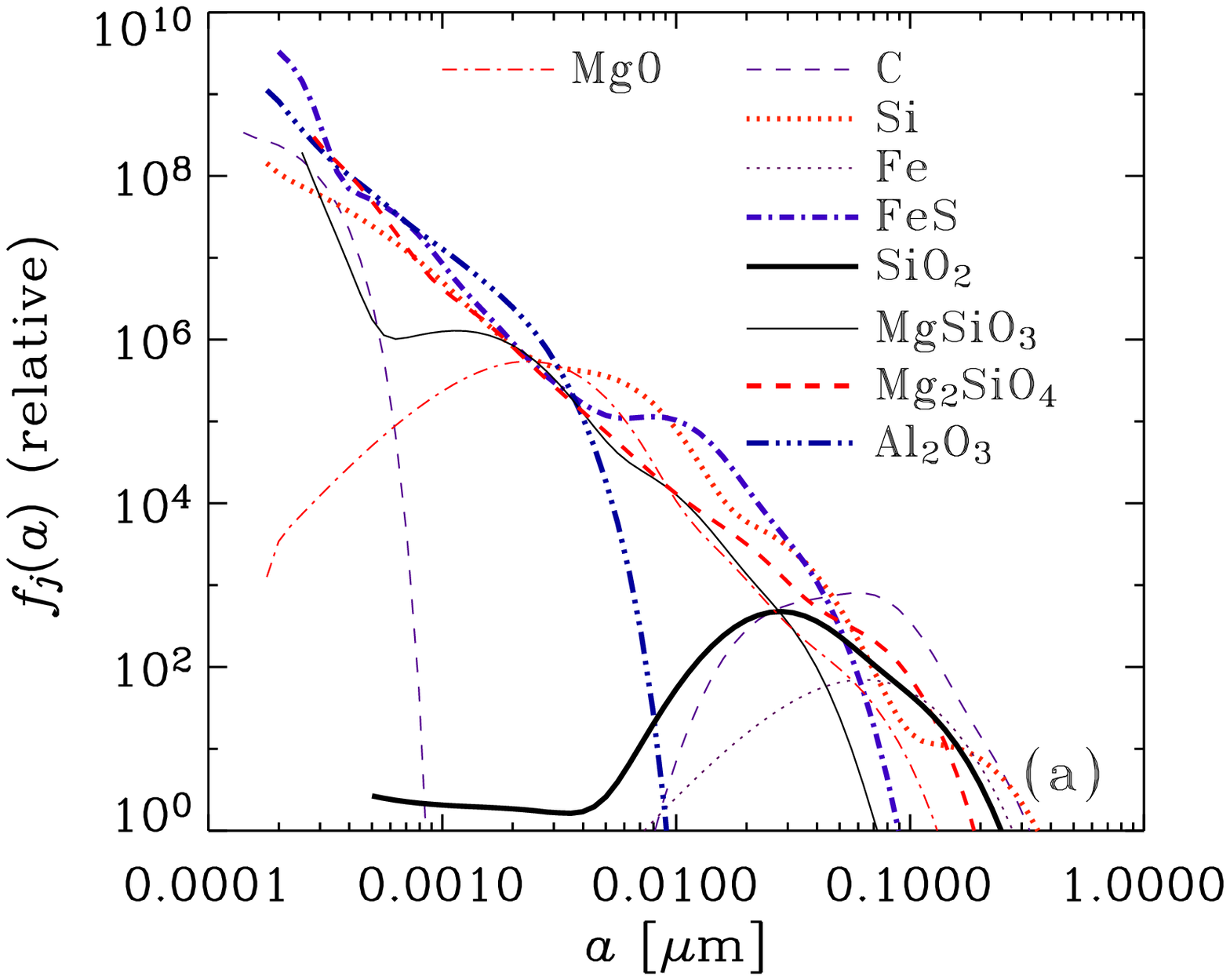}
\includegraphics[width=8cm]{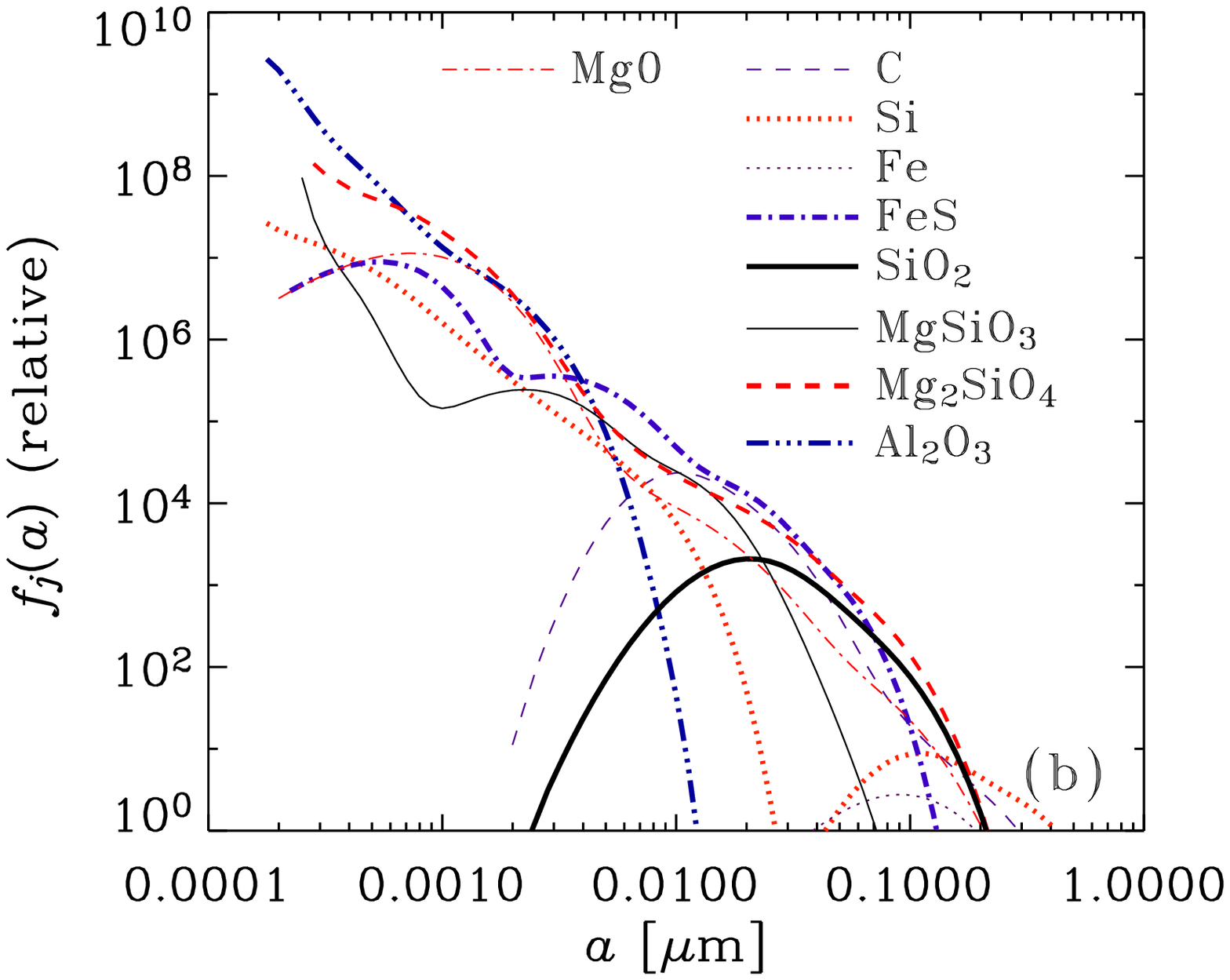}
\includegraphics[width=8cm]{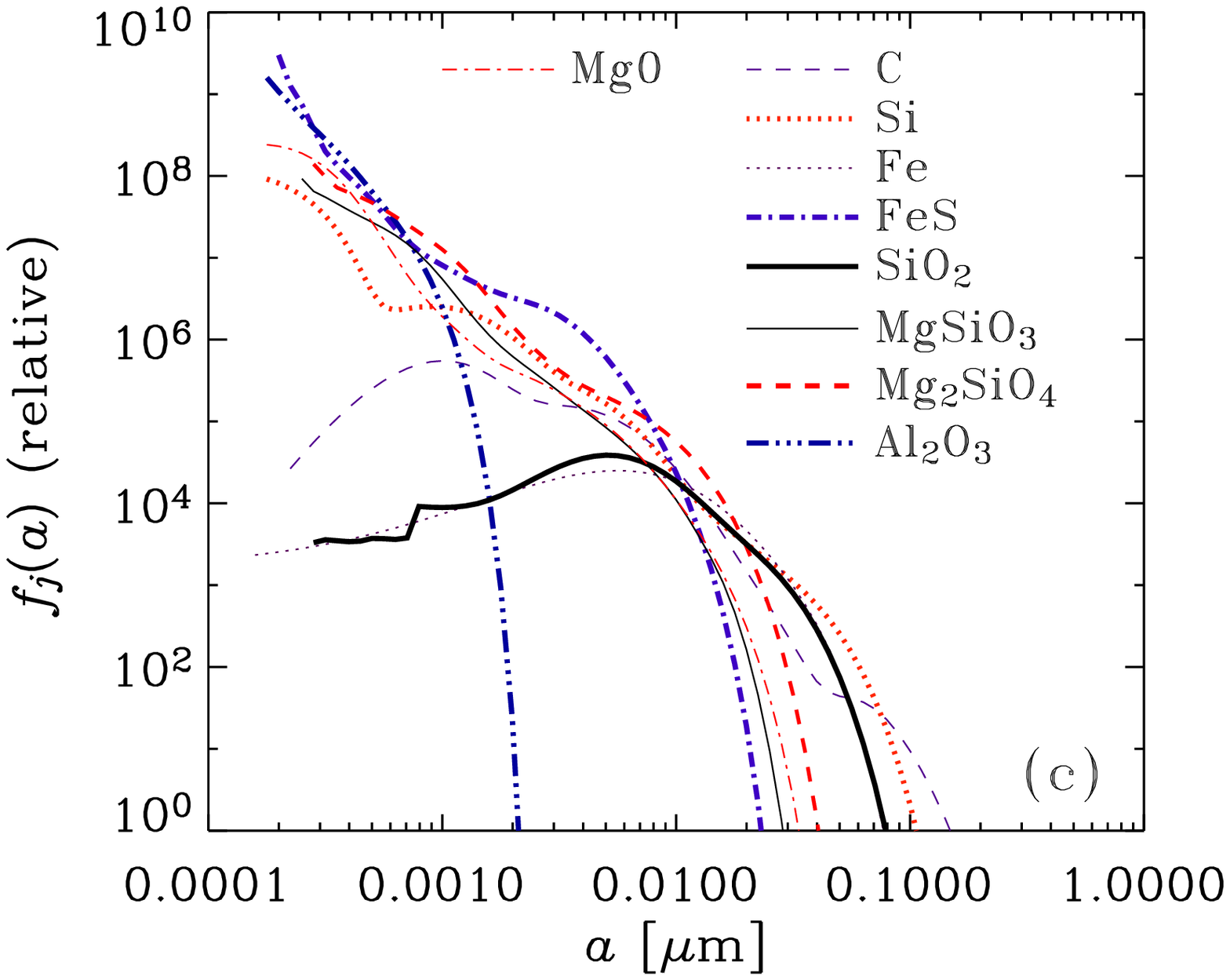}
\includegraphics[width=8cm]{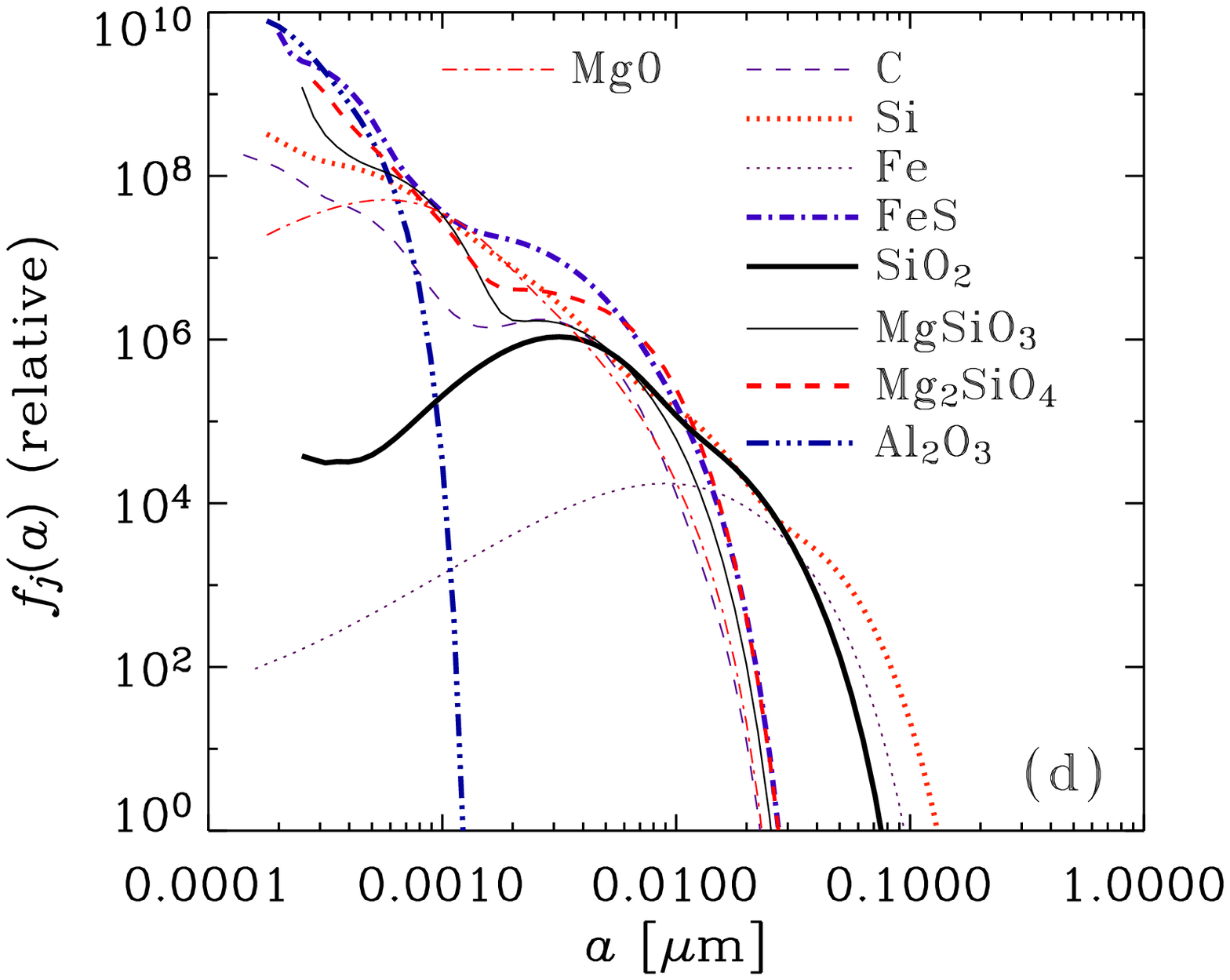}
\caption{Size distribution function of each grain species
in the unmixed ejecta with the progenitor masses of
(a) 13 $M_\odot$, (b) 20 $M_\odot$, (c) 25 $M_\odot$, and
(d) 30 $M_\odot$. The correspondence between the species and
lines is shown in the figures.
\label{fig:size_unmix}}
\end{figure*}

\begin{figure*}
\includegraphics[width=8cm]{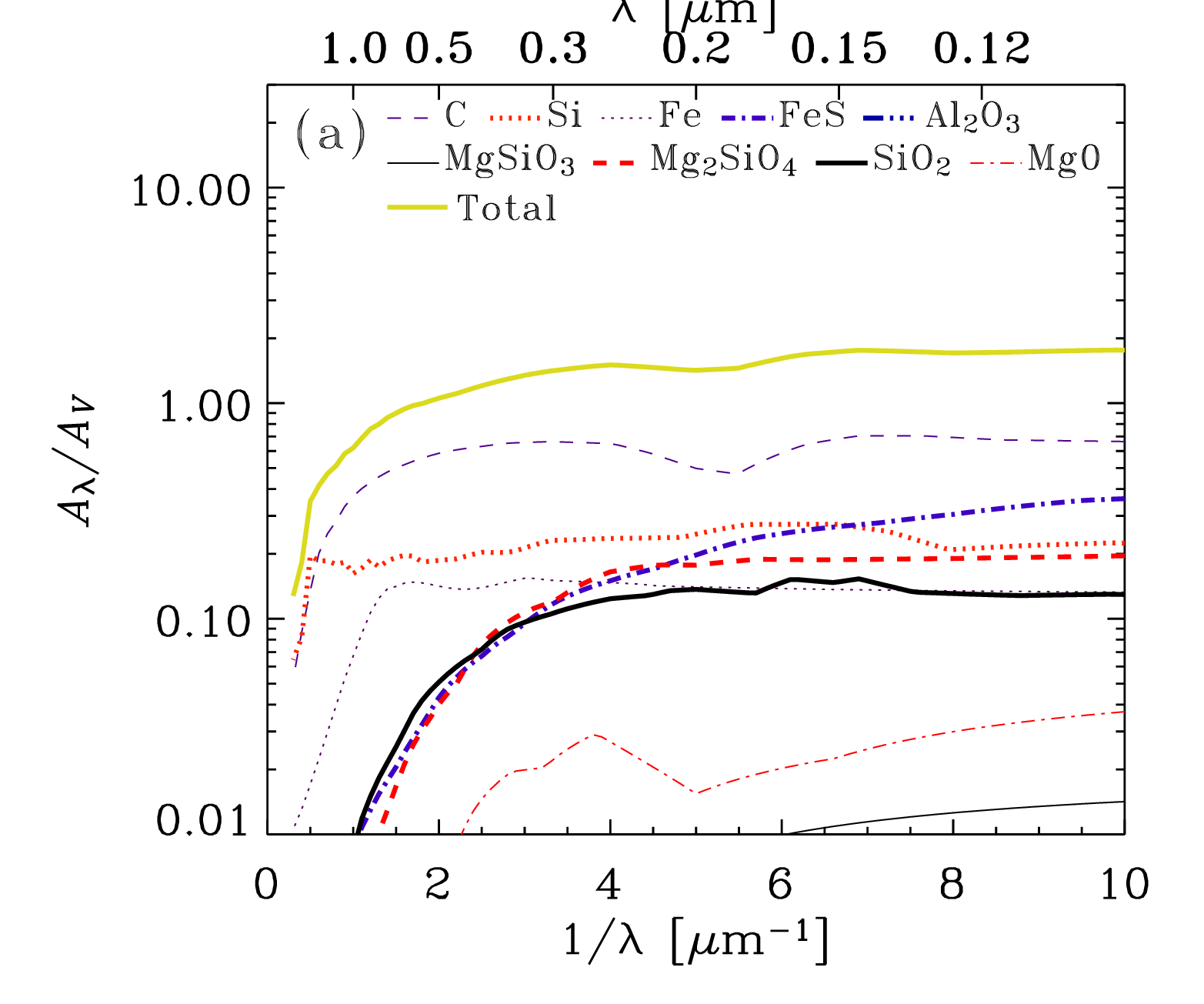}
\includegraphics[width=8cm]{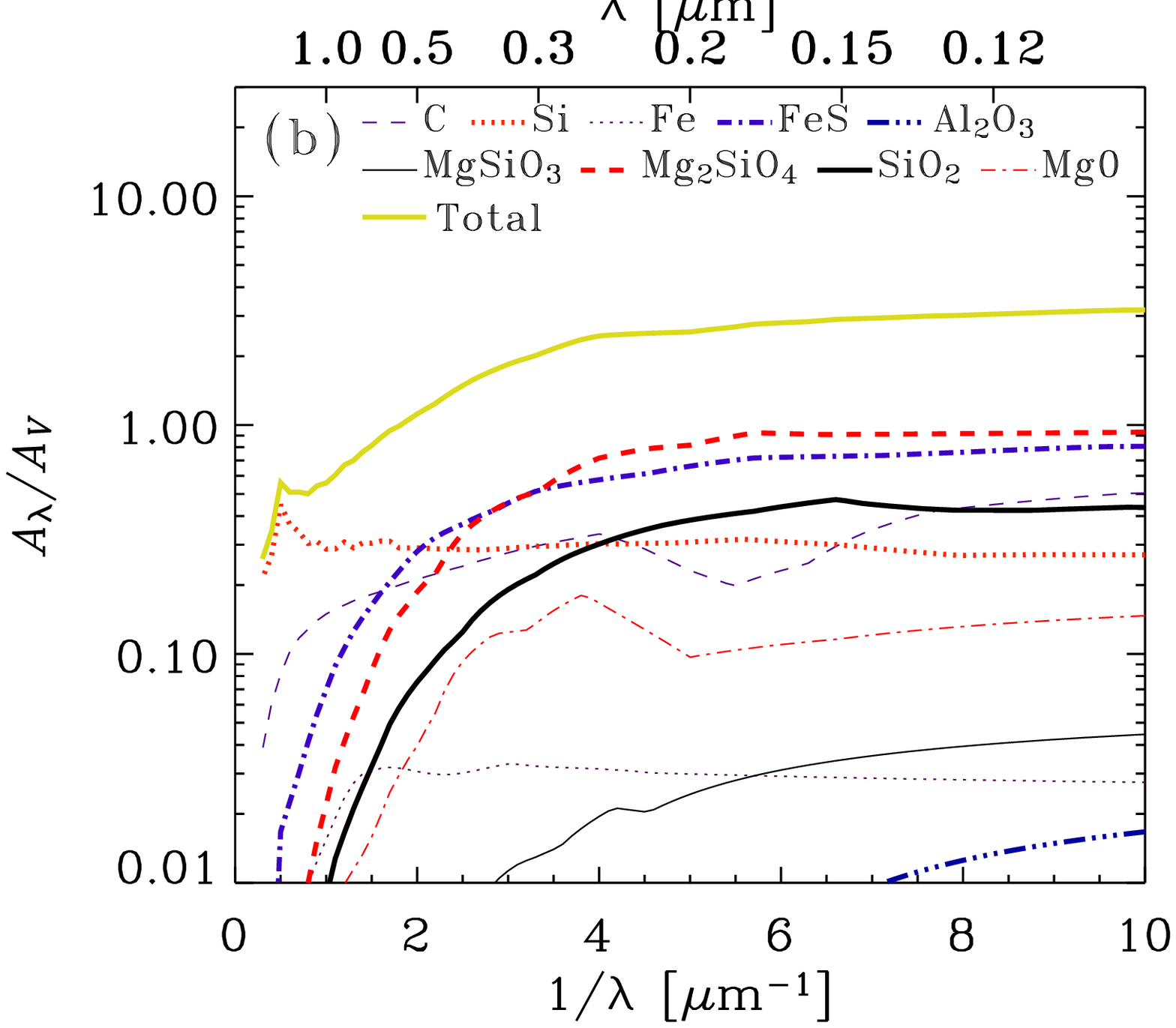}
\includegraphics[width=8cm]{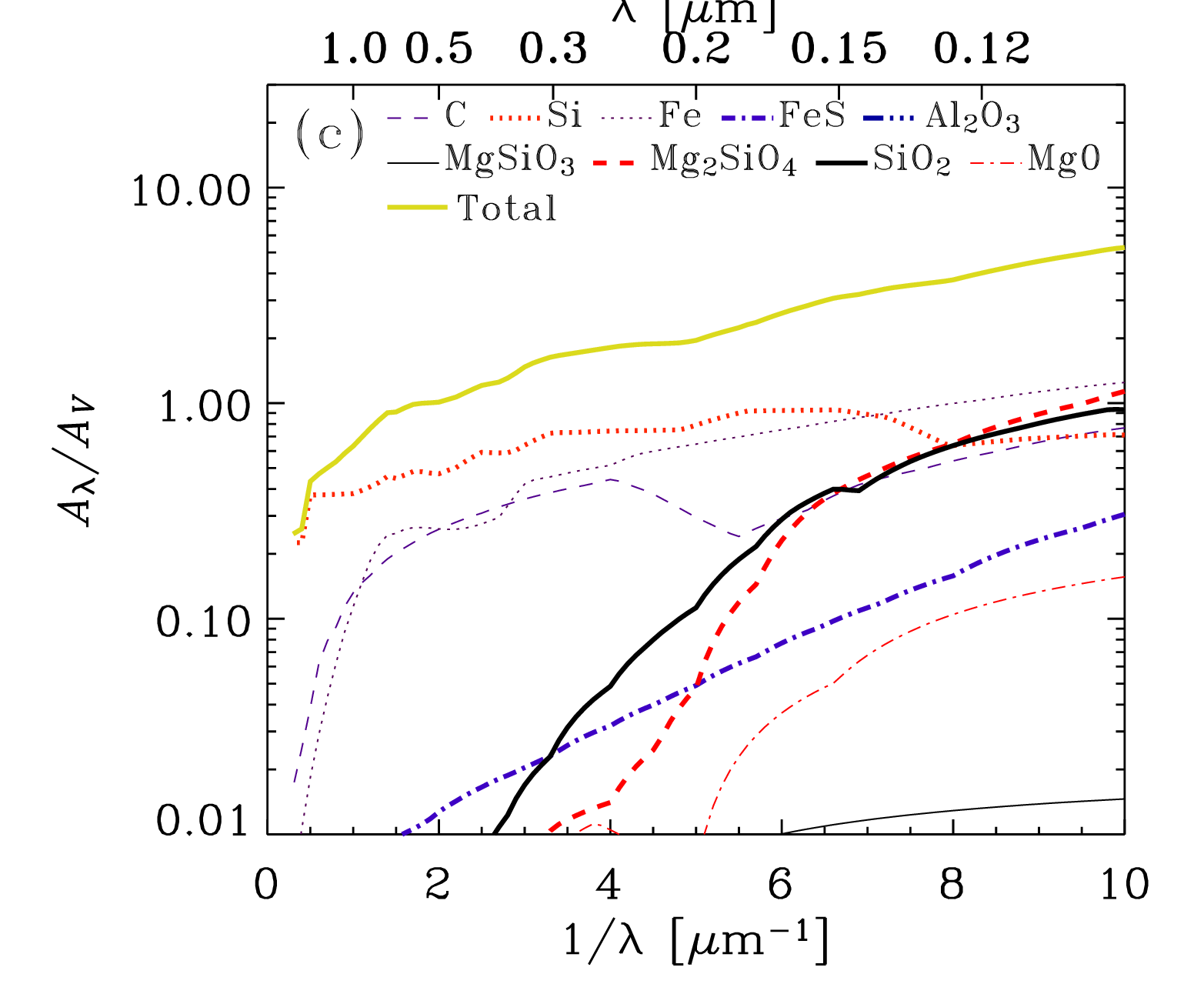}
\includegraphics[width=8cm]{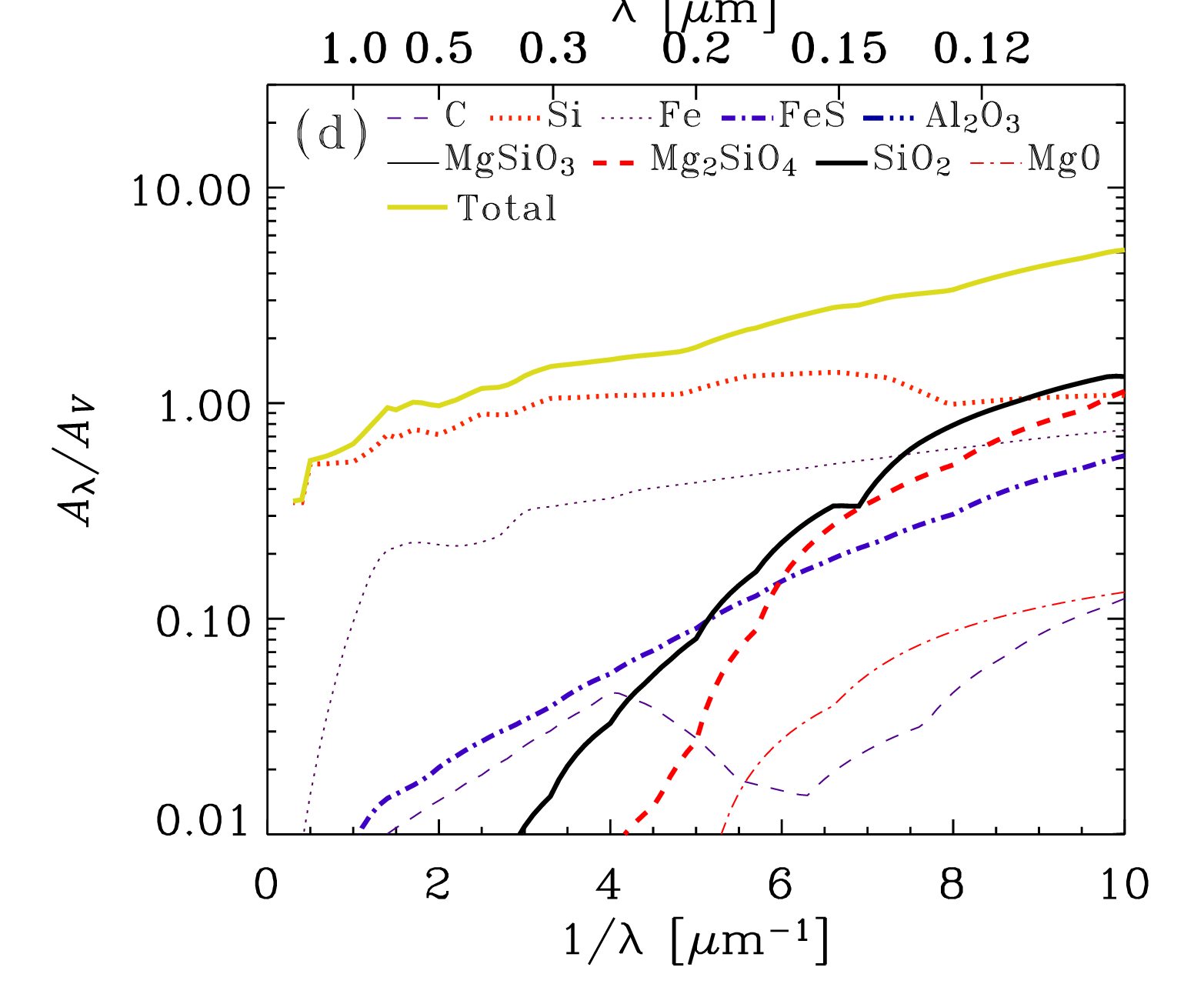}
\caption{Extinction curves calculated for the dust
production models of unmixed SN II ejecta with the
progenitor masses of (a) 13 $M_\odot$, (b) 20 $M_\odot$,
(c) 25 $M_\odot$, and (d) 30 $M_\odot$. Each extinction
curve is normalised
to the value at the $V$ band ($\lambda =0.55~\mu$m) of
the total extinction curve.
The contribution of each species is also
shown. The correspondence between the species and
lines is shown in the figures.
\label{fig:ext_unmix}}
\end{figure*}

\begin{figure*}
\includegraphics[width=8cm]{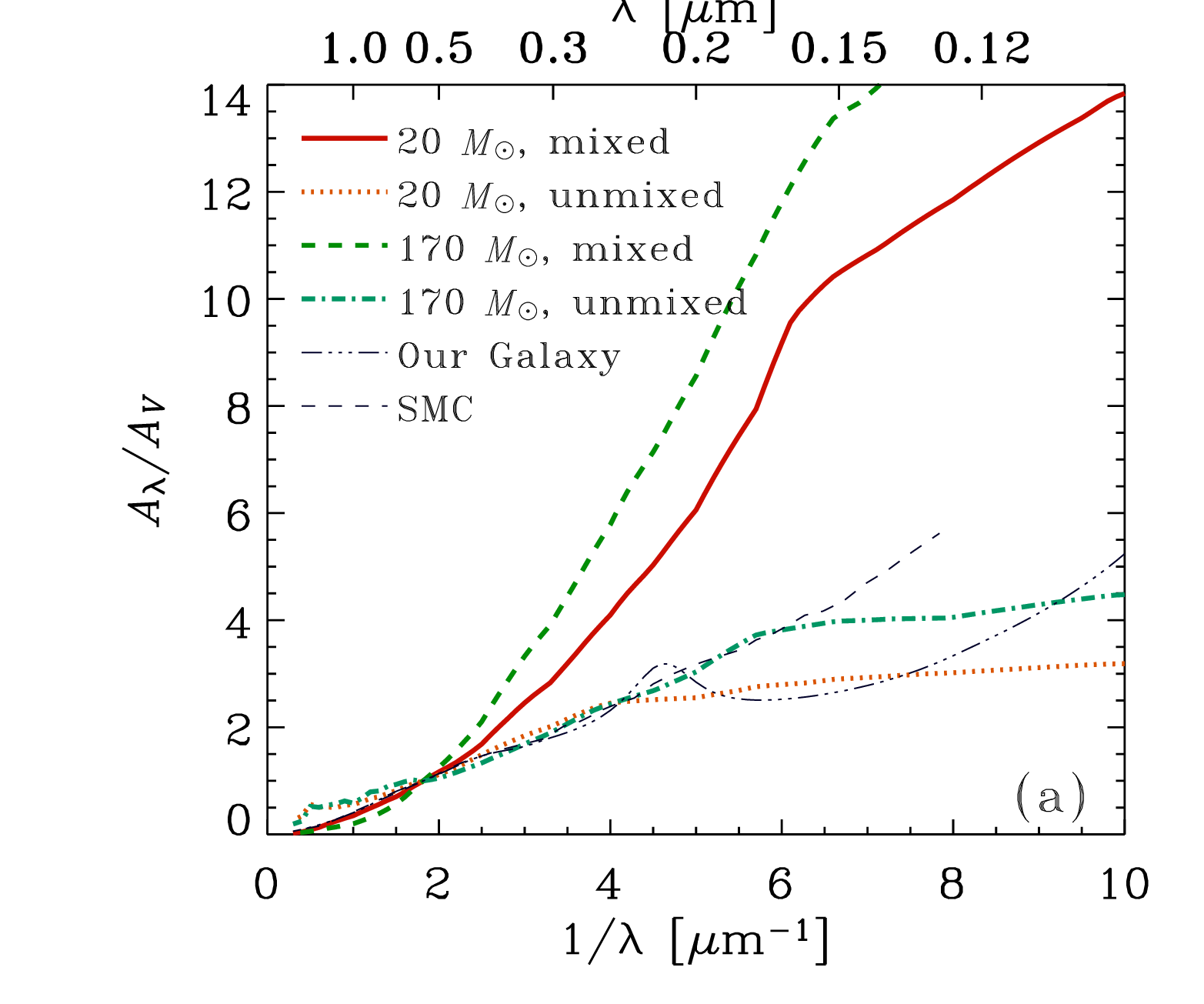}
\includegraphics[width=8cm]{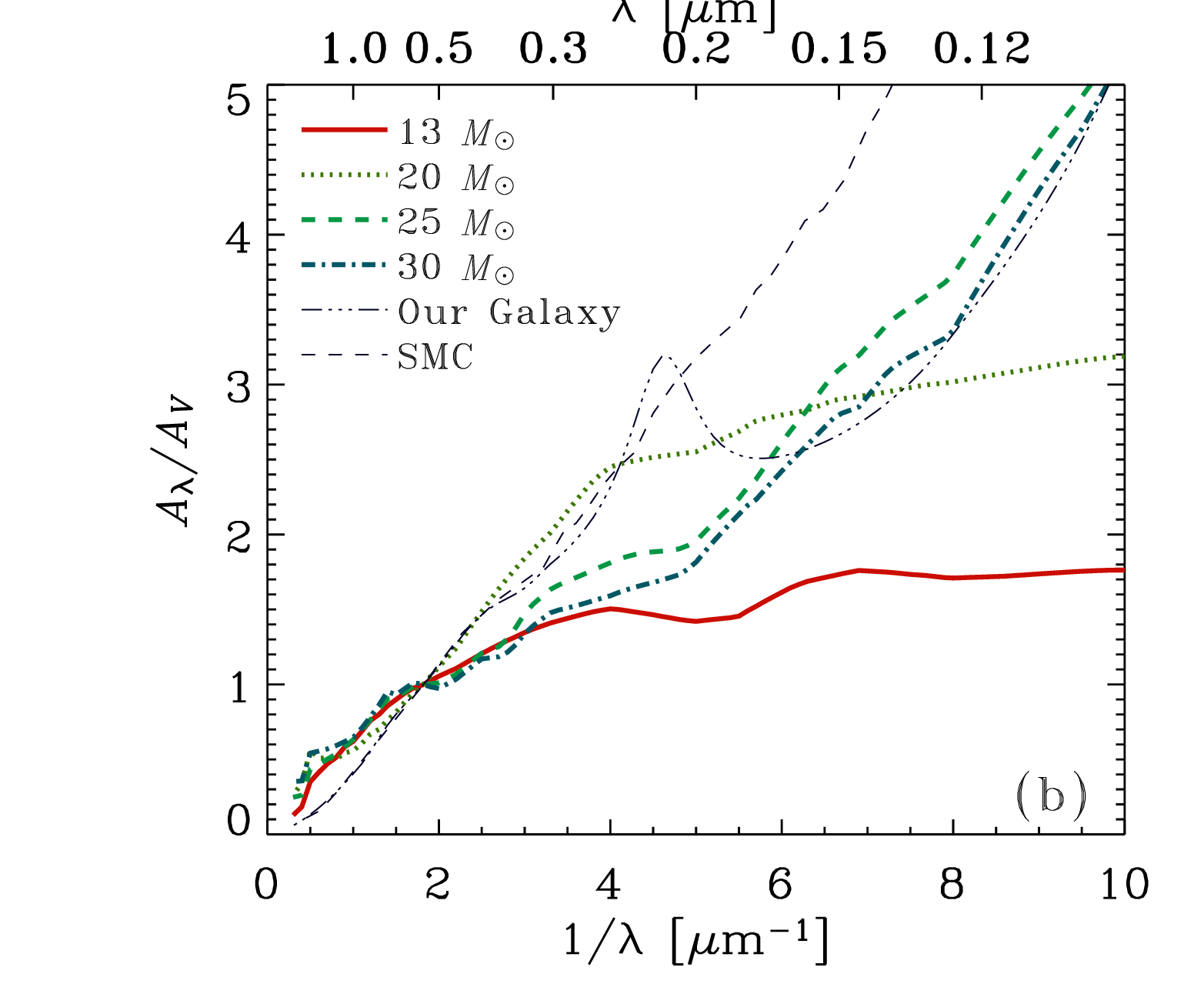}
\caption{Comparison between the extinction curves
calculated by our models. (a) Extinction curves of
various types of progenitors. The thick solid,
dotted, dashed, and dot-dashed lines correspond
to Models a, b, c, and d in Table \ref{tab:cases},
respectively. (b) Extinction curves for various
progenitor masses of unmixed Type II SNe. The
thick solid, dotted, dashed, and dot-dashed lines
correspond to the progenitor masses of 13, 20, 25,
and 30 $M_\odot$, respectively. We also show the
extinction curves of
the Galaxy (dot-dot-dot-dashed line) and the Small
Magellanic Cloud (SMC) (thin dashed line) only for
the reference.
It is not necessary that our model explains the
Galactic or SMC curve.
\label{fig:comparison}}
\end{figure*}

\begin{figure*}
\includegraphics[width=8cm]{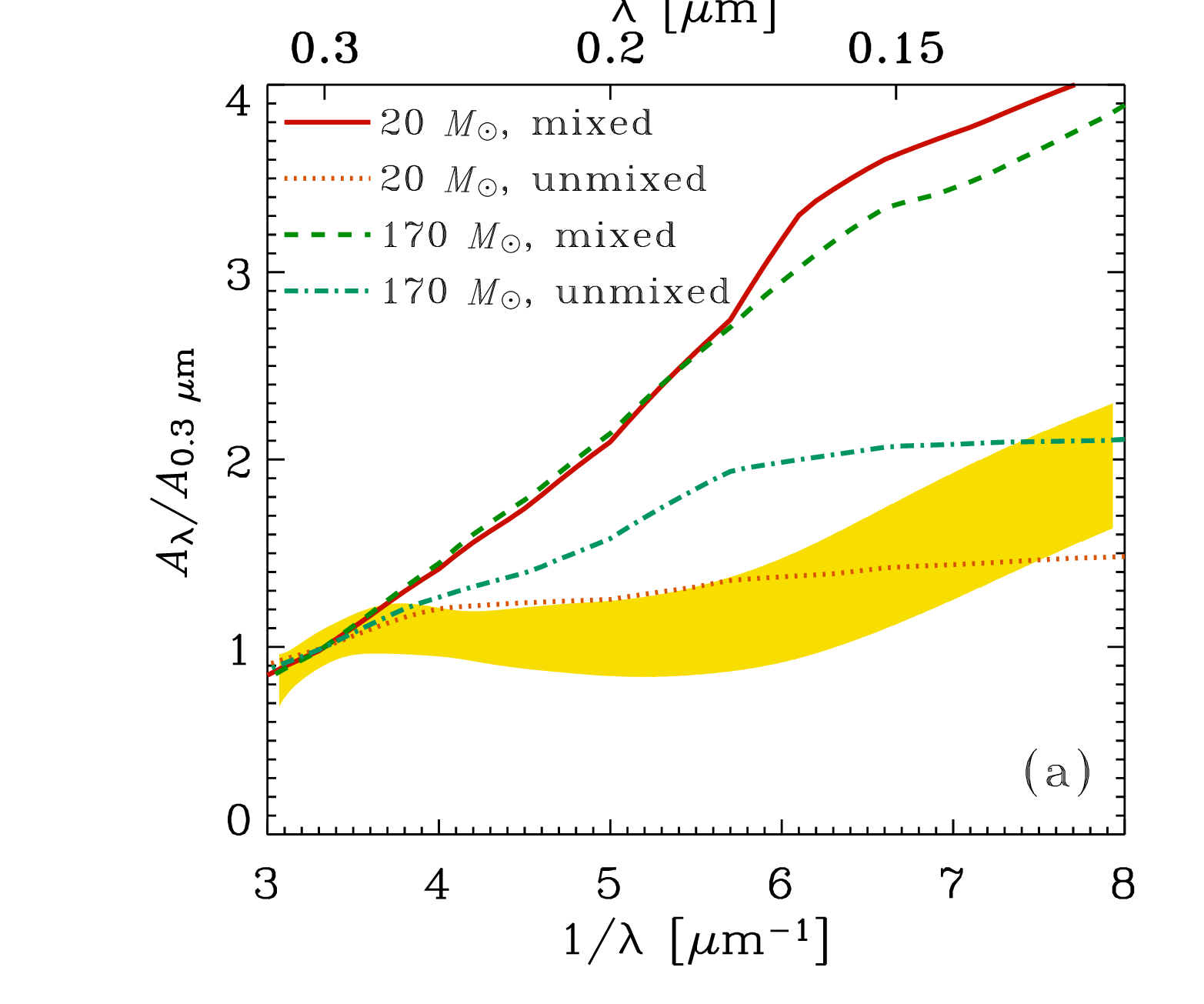}
\includegraphics[width=8cm]{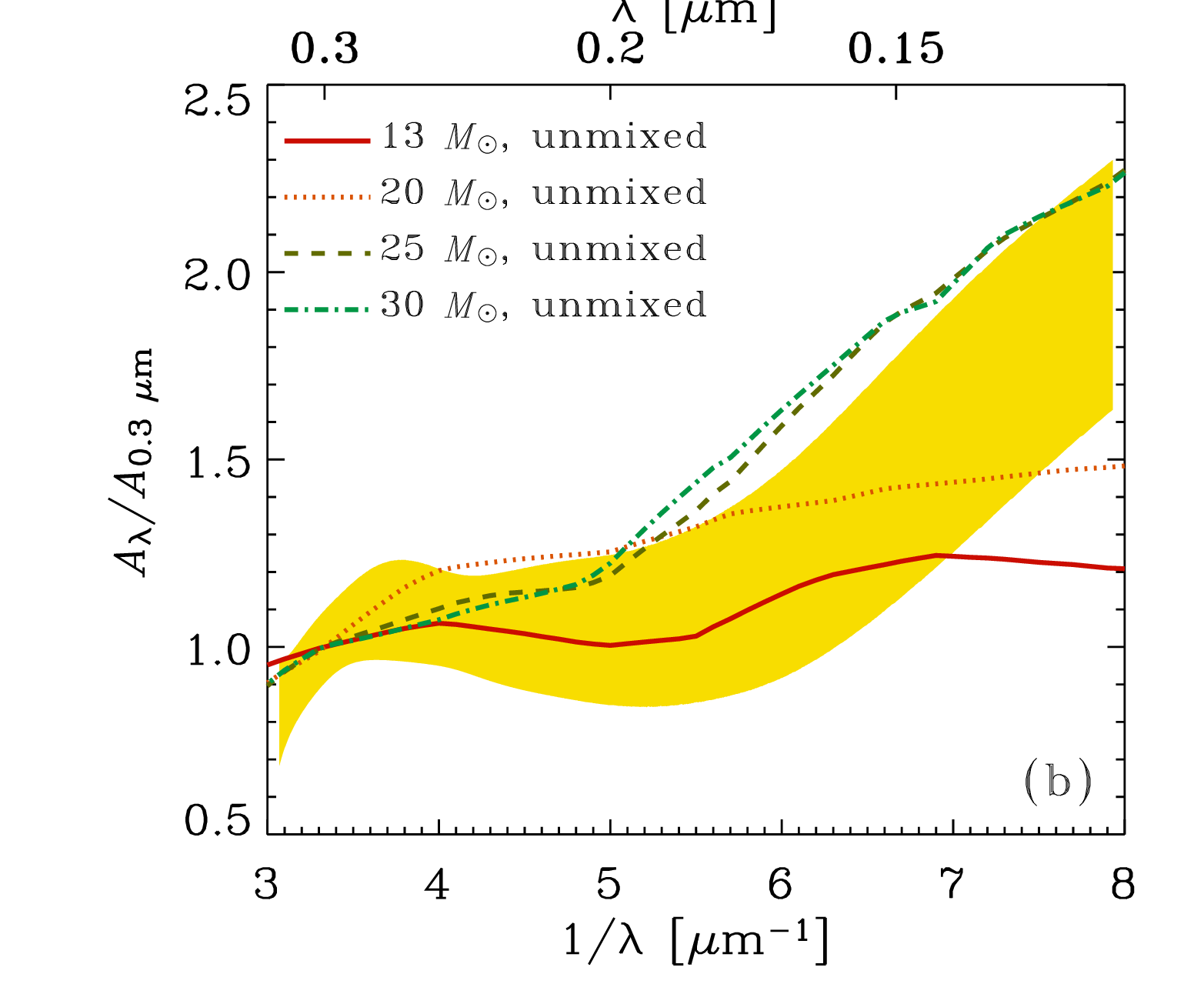}
\includegraphics[width=8cm]{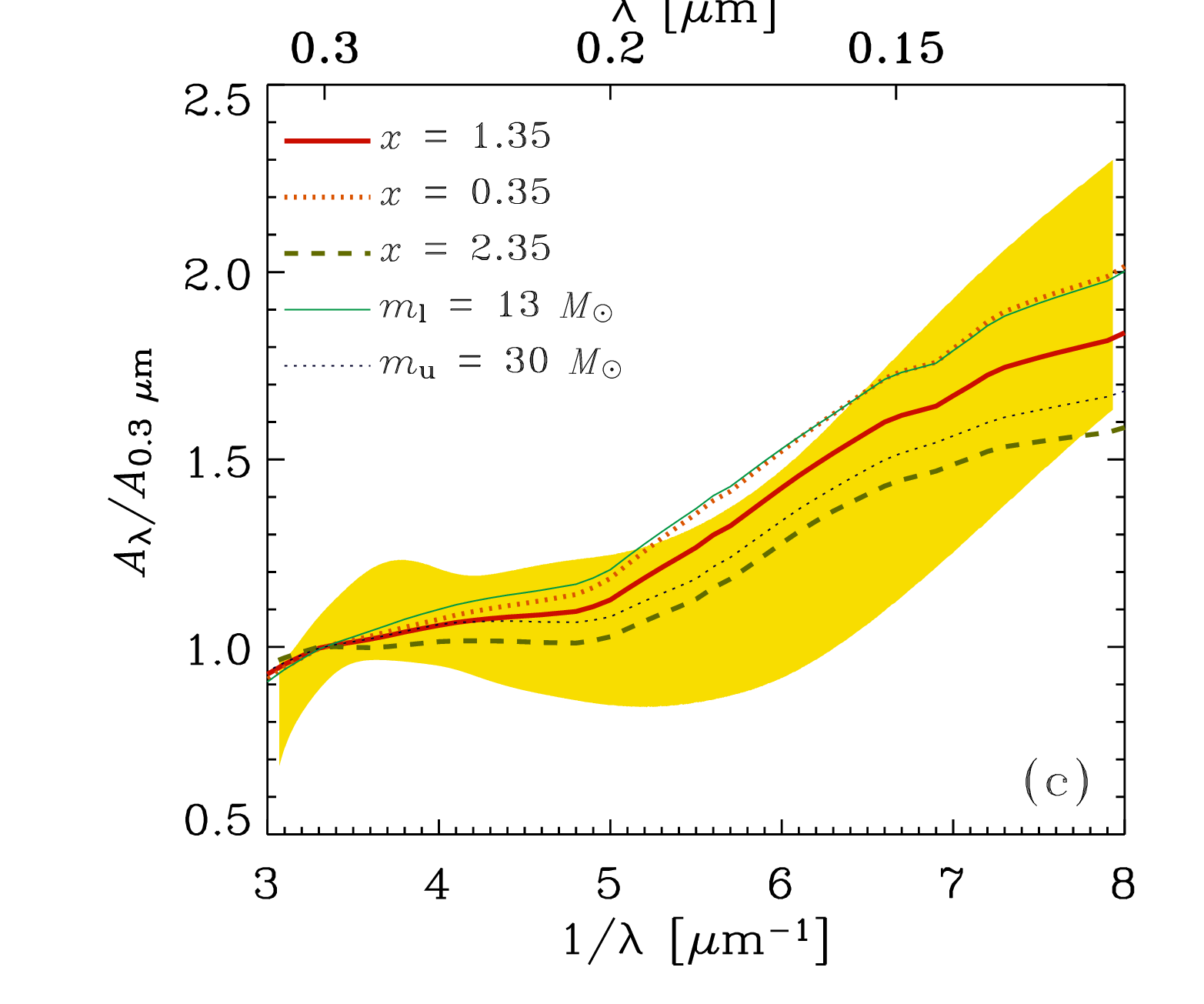}
\caption{Ultraviolet extinction curves normalised to
the extinction at 0.3 $\mu$m. The range
observationally derived by Maiolino et al.\ (2004b)
for \sdss\ is shown by the shaded area in each
figure. (a) Model predictions for various supernova
models (Models a--d in Table \ref{tab:cases}).
(b) Results for various progenitor masses unmixed
SNe II. The solid, dotted, dashed, and dot-dashed
lines correspond to the
progenitor masses of 13, 20, 25, and 30 $M_\odot$,
respectively. (c) Averaged extinction curves of unmixed
SNe II. The weight of each progenitor mass is
determined by initial mass functions, which are
parameterised by the power-law index ($x$) and
the lower and upper stellar masses ($m_{\rm l}$ and
$m_{\rm u}$, respectively).
The thick solid, dotted, and dashed lines represent
the results for $x=1.35$, 0.35, and 2.35,
respectively, with
$(m_{\rm l},\, m_{\rm u})=(8~M_\odot ,\, 40~M_\odot)$
(Models A, B, and C in Table \ref{tab:imf},
respectively).
The thin solid and dotted
lines show the results for
$(m_{\rm l},\, m_{\rm u})=(13~M_\odot ,\, 40~M_\odot)$,
and $(8~M_\odot ,\, 30~M_\odot)$,
respectively, with $x=1.35$ (Models D and E in
Table \ref{tab:imf}, respectively).
\label{fig:maiolino}}
\end{figure*}

The extinction curves of dust produced by the mixed SNe II
and PISNe are dominated by SiO$_2$
(Figures \ref{fig:theoretical}a and c). Actually, SiO$_2$
is the most abundant dust component in a mixed
20 $M_\odot$ SN II. However, in a 30 $M_\odot$ SN II, the
production of Mg$_2$SiO$_4$ is enhanced by 2.7 times
relatively to that of SiO$_2$, and in this case, the
contribution of Mg$_2$SiO$_4$ to the extinction curve
becomes 2.7 times larger. As a result, the extinction
curve is dominated by
the steep rising curve of Mg$_2$SiO$_4$ for
$\lambda\la 0.14~\mu$m.
For the mixed 170 $M_\odot$ PISN, the extinction curve is
also dominated
by SiO$_2$. Nozawa et al.\ (2003) also examine a larger
progenitor mass (200 $M_\odot$), where SiO$_2$ is the
dominant species in dust mass. Therefore, we expect
that the extinction curve of mixed PISNe always show
a curve similar to the one in
Figure \ref{fig:theoretical}c.

The extinction curve of dust produced by the unmixed
SNe II (Figure \ref{fig:theoretical}b) is dominated by
Mg$_2$SiO$_4$ and FeS for $\lambda\la 0.5~\mu$m. At
the longer wavelength, ($\lambda\ga 0.5~\mu$m), Si
dominates the curve. The extinction curves of unmixed
SNe II are flatter than those of mixed SNe II, because
a significant amount of large $(a\ga 0.1~\mu{\rm m})$
Si grains efficiently absorb long-wavelength photons.
In a more
massive SN II with the progenitor mass of 30 $M_\odot$,
the production of Mg$_2$SiO$_4$ is
enhanced by 2.2 times compared with Si, while the
amount of FeS is almost
the same; consequently the contribution
of Mg$_2$SiO$_4$ to the extinction curve is enhanced
by a factor of 2.2 relative to the Si contribution,
and the extinction curve becomes
steeper. On
the other hand, in less
massive unmixed SNe, the production of amorphous carbon
is enhanced, leading to the carbon-dominated flat
extinction curve. Therefore, the
extinction curve of unmixed SNe II
is sensitive to the progenitor mass, and the
progenitor mass dependence is shown in
Section \ref{subsec:unmixed}.

The extinction curve of unmixed PISNe is also flat,
because of Si contribution. The curve mildly rises
toward ultraviolet (UV) because of contribution from
Mg$_2$SiO$_4$. If the progenitor mass is
200 $M_\odot$, the production of Si is enhanced by
twice relative to Mg$_2$SiO$_4$, and the
contribution of Si becomes comparable to that of
Mg$_2$SiO$_4$ even in UV. This makes the
extinction curve flatter than that shown in
Figure \ref{fig:theoretical}d.

In general, the extinction curves of mixed cases are
steeper than that of unmixed cases.
This is because the dust opacity of the mixed SNe II
is dominated by small
($a\sim 0.01~\mu$m) SiO$_2$ grains. On the
contrary, large Si grains produced in unmixed SNe
have a large cross section up to the near infrared
(NIR).

All the four extinction curves are compared in
Figure \ref{fig:comparison}a. We also show the
extinction curves of the Galaxy
(Cardelli, Clayton, \& Mathis 1989 with $R_V=3.1$)
and the Small Magellanic Cloud (SMC) (Pei 1992).
It is not necessary that our theoretical curves
explain the Galactic and SMC curves, since in those
two environments, dust grains are also produced by
late-type stars with long lifetimes.

We do not enter deeply the infrared regime.
The infrared properties of dust grains are addressed
in Takeuchi et al.\ (2004, in preparation;
see also Takeuchi et al.\ 2003).

\subsection{Progenitor mass dependence of unmixed
SNe II}\label{subsec:unmixed}

As mentioned in Section \ref{subsec:theor}, the
progenitor mass dependence is significant in the
case of unmixed SNe II. Therefore, we examine
various progenitor masses examined by
Nozawa et al.\ (2003); i.e.\ the progenitor
masses of 13, 20, 25, and 30 $M_\odot$. In
Figure \ref{fig:size_unmix}, we present the
size distribution for each species
(Figures \ref{fig:size_unmix}a, b, c, and d
correspond to the progenitor masses of 13,
20, 25, and 30 $M_\odot$, respectively).
Based on those size distributions, we calculate
the extinction curves by the method described
in Section \ref{subsec:method}. The extinction
curve as well as the contribution from each
species is shown in Figure \ref{fig:ext_unmix}.
We see that the 13 $M_\odot$ extinction curve
is dominated by carbon grains, whose typical
size is large ($\sim 0.1~\mu$m). Such large
carbon grains produce a flat extinction curve
as presented in
Figure \ref{fig:ext_unmix}a. As the progenitor
mass becomes larger, the contribution from
Si becomes larger. Although the Si extinction
curve is flat, other species
such as Mg$_2$SiO$_4$, FeS, Fe, and SiO$_2$
contribute to the rising curve at short
wavelengths in a complex way. As a result, if
the dust production occurs in the unmixed
SNe II, the extinction curve tends to be
steeper as the progenitor mass becomes
larger. The four extinction curves are compared
in Figure \ref{fig:comparison}b.

\subsection{Comparison with standard silicate and
graphite}

It is useful to compare our prediction with a standard
``astronomical silicate'' in Draine \& Lee (1984).
Their optical constants in UV are based on olivine
(Mg, Fe)$_2$SiO$_4$ (Huffman \& Stapp 1973), and the
nearest species in our model is Mg$_2$SiO$_4$.
Our optical constant assumed for Mg$_2$SiO$_4$ is almost
the same as that of ``astronomical silicate'' in UV, but
the astronomical silicates have a larger cross section
in the optical and NIR than our Mg$_2$SiO$_4$.
We calculate the
extinction curve of the astronomical silicate
by using the optical constant of Draine \& Lee (1984)
and the size distribution of Mg$_2$SiO$_4$ calculated
by Nozawa et al.\ (2003).
The extinction curve calculated by this method is
the same as our extinction curve of Mg$_2$SiO$_4$ in
UV, but at the $V$ band, the difference becomes
at most a factor of $\sim 2$. Then the astronomical
silicate predicts an extinction curves with a
shallower slope in
NIR. However, the difference does not affect our
results, since the contribution of Mg$_2$SiO$_4$ is
important in UV. The Mg$_2$SiO$_4$ data
presented in Scott \& Duley (1996) has a similar optical
constants to Draine \& Lee (1984) in UV and to
our adopted values in optical and NIR.

The contribution of carbon grains is small in the four
cases presented in Figure \ref{fig:theoretical}.
If we use the optical constant of graphite in
Draine \& Lee (1984) and the size distribution of
carbon grains in Nozawa et al.\ (2003), we see a weak
bump around 2200 \AA\ and the overall contribution of
carbon is reduced by a factor of two.
The change of carbon
optical properties does not affect the total extinction.
However, Nozawa et al.\ (2003) show that carbon becomes
a principal species if the progenitor mass is around
$13~M_\odot$. In particular, the
production of Mg$_2$SiO$_4$ is much reduced in a
13 $M_\odot$ SN II. Indeed, carbon grains dominate the
extinction curve in
13 $M_\odot$ unmixed SNe. Regardless of which
optical constant we adopt for carbon grains, we obtain
a flat UV extinction curve for a 13 $M_\odot$
unmixed SN II (see Section \ref{subsec:comp}).

\subsection{Useful quantities}\label{subsec:useful}

The extinction curve is often characterised by the
parameter $R_V$ defined as
\begin{eqnarray}
R_V\equiv\frac{A_V}{E(B-V)}\, ,
\end{eqnarray}
where $E(B-V)\equiv A_B-A_V$, and $B$ indicates
the $B$-band wavelength (0.44 $\mu$m). $R_V$
roughly quantify the inclination of the extinction
curve in the optical. The value calculated for each
theoretical extinction curve
is shown in Table \ref{tab:cases}.

It is also useful to relate the extinction to
the dust amount, because dust mass also constrain
the dust production model in galaxies (e.g.\
Hirashita \& Ferrara 2002). For this aim, we should
determine the normalisation constant
$C$ in equation (\ref{eq:tau}). 
The constant $C$ is determined
to realise the total dust column density as
\begin{eqnarray}
\mu m_{\rm H}N_{\rm H}{\cal D}=C\sum_j
\int_0^\infty\frac{4}{3}\pi a^3\delta_jf_j(a)\,
{\rm d}a\, ,\label{eq:norm}
\end{eqnarray}
where $\mu$ is the gas mass per hydrogen nucleus
(assumed to be 1.4 in this paper; Spitzer 1978),
$m_{\rm H}$ is the mass of a hydrogen atom,
$N_{\rm H}$ is the column density of hydrogen
nuclei, ${\cal D}$ is the dust-to-gas mass ratio,
and $\delta_j$ is the material density of grain
species $j$. We calculate $\delta_j$ from the
radius per unit molecule listed in
Robie \& Waldbaum (1968) and the mass of
atoms averaged with the isotope ratio in SN
ejecta. The material density $\delta_j$ for each
species is listed in Table \ref{tab:species}.

To relate the dust cross section to the dust mass,
we define the cross section per unit dust mass,
$\langle\sigma_{\rm d}(\lambda)/m_{\rm d}\rangle$, as
\begin{eqnarray}
\langle\sigma_{\rm d}(\lambda )/m_{\rm d}\rangle
\equiv\frac{\displaystyle\sum_j\int_0^\infty\pi a^2
Q_{{\rm ext},\, j}(\lambda ,\,a)f_j(a)\, {\rm d}a}
{\displaystyle\sum_j\int_0^\infty\frac{4}{3}\pi
a^3\delta_jf_j(a)\, {\rm d}a}\, .\label{eq:cross}
\end{eqnarray}
Then, the following expression of extinction
is derived by using equations (\ref{eq:tau}),
(\ref{eq:extj}), (\ref{eq:extall}),
(\ref{eq:norm}), and (\ref{eq:cross}):
\begin{eqnarray}
A_\lambda =1.086\mu m_{\rm H}N_{\rm H}{\cal D}
\langle\sigma_{\rm d}(\lambda)/m_{\rm d}\rangle\, .
\label{eq:AvsN}
\end{eqnarray}
We provide the values of
$\langle\sigma_{\rm d}(V)/m_{\rm d}\rangle$
(the cross section in the $V$ band per unit dust
mass) in Table \ref{tab:cases}. By using this value
the dust amount can be quantified if we know
$N_{\rm H}$ from other observations (for
example, observations of Ly$\alpha$ absorption
line). Then, $A_V$ is obtained
observationally for primeval galaxies
(a system in which dust is predominantly supplied
by SNe II or PISNe), we obtain the dust-to-gas ratio.
The colour excess $E(B-V)$ may be more easily
obtained, and in this case, $R_V$ listed in
Table \ref{tab:cases} could be used to derive
the extinction $A_V$.

In the Galactic ISM,
$N_{\rm H}/A_V=1.9\times 10^{21}~{\rm cm}^{-2}~
{\rm mag}^{-1}$ and ${\cal D}=6\times 10^{-3}$
(Spitzer 1978). By using equation (\ref{eq:AvsN}),
we obtain
$\langle\sigma_{\rm d}(V)/m_{\rm d}\rangle =
3.4\times 10^4~{\rm cm}^2~{\rm g}^{-1}$ for
the Galactic dust. We have obtained similar values
in Models b and d (both assume unmixed SNe) and
significantly smaller ones in Models a and c
(both assume mixed SNe).

The dust cross section per mass is also calculated
for a UV wavelength (0.3 $\mu$m). In
Table \ref{tab:cases}, we list
$\langle\sigma_{\rm d}(0.3~\mu{\rm m})/m_{\rm d}\rangle$.
The difference between the models is relatively small.
Therefore, the UV dust extinction is a better tracer
of the dust column density than the optical
extinction.

\section{OBSERVATIONAL DISCUSSION}\label{sec:obs}

\subsection{Comparison with high-$z$ data}
\label{subsec:comp}

At present, there are few observational works of the
extinction curves at $z>5$, where the cosmic age is
shorter than 1 Gyr. With this short timescale, the
dust is predominantly formed by SNe II and PISNe,
since their progenitors have short lifetimes
while evolved low mass stars require longer
timescales to evolve and produce dust.
Maiolino et al.\ (2004b) use the extinction curve of
\sdss\ to test the hypothesis that dust is
predominantly supplied by SNe II. They
explain the extinction curve by the
dust production model of Todini \& Ferrara (2001).
They also investigate various initial stellar
metallicities from 0 to solar
after averaging the grain properties for
different SNe II over the Salpeter stellar initial
mass function (IMF),
and find that their theoretical extinction curves
agree with the observational data of \sdss\
in the whole metallicity range.

We also use the restframe UV extinction curve of
\sdss\ shown in Figure 2 of
Maiolino et al.\ (2004b). The
extinction curve is flat at $\lambda >0.17~\mu$m,
and it increases with a smaller rate than the SMC
extinction curve toward the shorter wavelength
at $\lambda <0.17~\mu$m. The plausible range derived
by Maiolino et al.\ (2004b) is
shown by the shaded areas in Figure \ref{fig:maiolino}.
In Figure \ref{fig:maiolino}a, the
four theoretical curves calculated by our model are
shown. The extinction curves calculated with the
mixed SN models are too steep to explain the
observational data.

The models with the unmixed SNe agrees quite well
with the observational data. The extinction curve of
unmixed SNe II depends on the progenitor mass as
shown in Section \ref{subsec:unmixed}. In
Figure \ref{fig:maiolino}b, we show the UV
extinction curves of unmixed SNe II with the
progenitor masses of 13, 20, 25, and 30 $M_\odot$
(solid, dotted,
dashed, and dot-dashed lines, respectively).
The flat behaviour of
13 $M_\odot$ extinction curve comes from the
carbon contribution, which produces a slight bump
around $1/\lambda\sim 4~\mu{\rm m}^{-1}$. The
rise toward the shorter wavelength is mainly caused
by Mg$_2$SiO$_4$. Since the production of
Mg$_2$SiO$_4$ is enhanced in the massive progenitors,
the extinction curve becomes steeper for more
massive progenitors. The unmixed SN II models are
roughly consistent with the current observational
data at high $z$. It is interesting to note that
the observational ranges lie between the flat
curve predicted for low-mass (13 $M_\odot$ and
20 $M_\odot$) progenitors and the
steep curve predicted by high-mass
(25 $M_\odot$ and 30 $M_\odot$) progenitors.
Therefore, the mixture of unmixed SNe II with various
progenitor mass may explain the observational
extinction curve.

Therefore, we calculate extinction curves weighted
by IMF:
\begin{eqnarray}
A_\lambda (\phi )\equiv\int_{m_{\rm l}}^{m_{\rm u}}
A_\lambda (m)\phi (m)\,{\rm d}m\, ,
\end{eqnarray}
where $m$ is the progenitor mass, $m_{\rm l}$ and
$m_{\rm u}$ are, respectively, the lower and upper
mass limits of stars which cause SNe II,
$A_\lambda (m)$ is the extinction calculated for
the progenitor mass $m$ (correctly weighted for
produced dust mass; i.e., $A_\lambda (m)$ is large
if the progenitor produces a large amount of dust),
and $\phi (m)$ is the IMF (the number of stars with
the mass range of $[m,\, m+{\rm d}m]$ is
proportional to $\phi (m)\,{\rm d}m$). The
normalisation of
$\phi (m)$ is not important in this paper, because
the extinction curve is always shown after being
normalised by $A_V$ and the normalising constant
of $\phi (m)$ is cancelled out. We assume the
following power-law form of the IMF:
\begin{eqnarray}
\phi (m)=Km^{-(x+1)}\, ,
\end{eqnarray}
where $K$ is the normalising constant. The Salpeter
IMF is reproduced by $x=1.35$. We examine
$x=1.35$, 0.35, and 2.35. An appropriate stellar
mass range for SNe II (core collapse SNe) is
selected as
$m_{\rm l}=8~M_\odot$ and $m_{\rm u}=40~M_\odot$
(Heger \& Woosley 2002). We use the calculated
extinction curves of unmixed SNe II with
$m=13$, 20, 25, 30 $M_\odot$, and interpolate
or extrapolate the values to obtain the
extinction curves of arbitrary progenitor mass.
The slopes $x=0.35$ (2.35) represents the case
where massive
(less massive) stars are selectively produced.
The extinction curves weighted for the IMFs are
shown in Figure \ref{fig:maiolino}c, where
the thick solid, dotted, and dashed lines represent
the results with $x=1.35$, 0.35, and 2.35,
respectively.
We also investigate the effect of varying
$m_{\rm l}$ and $m_{\rm u}$ with $x=1.35$: the thin
solid and dotted lines in Figure \ref{fig:maiolino}c
show the results with
$(m_{\rm l},\, m_{\rm u})=(13~M_\odot ,\, 40~M_\odot)$
and $(8~M_\odot ,\, 30~M_\odot)$, respectively.
All the examined IMFs are summarised in
Table \ref{tab:cases}, and are labeled as Models
A--E.

As expected from Figure \ref{fig:maiolino}b, the
contribution from massive SNe II tends to increase
the slope of the extinction curve. Then,
Models B and D predict steeper extinction curves
than the data of Maiolino et al.\ (2004b).
Therefore, the contribution from the SNe II whose
progenitor mass is around 13 $M_\odot$ is necessary
to obtain the flat extinction curve consistent
with the observational data. However, in order to
reproduce the rise toward the shorter wavelength
$\lambda\la 0.15~\mu$m, the contribution from
massive SNe II is necessary. In particular,
Model C predicts an extinction curve slightly
inconsistent around
$1/\lambda\sim 8~\mu{\rm m}^{-1}$.
We should emphasize that the Salpeter IMF
(Model A) reproduces the observed data very well.

\begin{table}
\centering
\begin{minipage}{80mm}
\caption{Initial mass functions.}
\begin{tabular}{@{}cccc@{}}\hline
Model & $x$ & $m_{\rm l}$ & $m_{\rm u}$ \\
 & & ($M_\odot$) & ($M_\odot$) \\
\hline
A & 1.35 & 8  & 40 \\
B & 0.35 & 8  & 40 \\
C & 2.35 & 8  & 40 \\
D & 1.35 & 13 & 40 \\
E & 1.35 & 8  & 30 \\
\hline
\end{tabular}
\label{tab:imf}
\end{minipage}
\end{table}

\subsection{Observational strategy}

A large sample of high-$z$ quasars and galaxies will
be obtained by future observations. Since the cosmic
age at $z=5$ is about 1 Gyr, the dust supplied by
late-type stars does not dominate the total dust
amount at $z>5$. Therefore, if future observations
collect a large spectroscopic sample of quasars
at $z>5$, we can directly investigate the extinction
curve of dust, whose source is probably SNe II
and/or PISNe. Although we have calculated
the dust formation based on the PopIII progenitors,
we can apply our model to metal-enriched systems as
long as the dust is predominantly formed by SNe II
and/or PISNe, because the dust composition and
size distribution is much less sensitive to the
progenitor metallicity than to the progenitor mass.

The shape of the extinction curve of high-$z$ galaxies
can be compared with the theoretical curves in this
paper to constrain the size and composition of grains.
The measurement of extinction enables us to measure the
dust-to-gas ratio of high-$z$ galaxies, if we use
equation (\ref{eq:AvsN}) and the cross section per
dust mass listed in Table \ref{tab:cases}.
Therefore, the dust production history in high-$z$
universe can also be investigated based on this
paper.

Some theoretical works (Bromm \& Larson 2004 and
references therein) suggest that first stars born from
primordial gas (PopIII stars) are massive. If the gas
metallicity is less than $\sim 10^{-5}Z_\odot$
($Z_\odot$ is the solar metallicity), massive stars
may selectively form
(Schneider et al.\ 2003; see also Omukai 2001),
causing PISNe at the end of their lives. Therefore,
the extinction curve of extremely metal-poor galaxies
could be compared with our curves calculated with
PISN models. The extinction curve of \sdss\
(Maiolino et al.\ 2004b) has been
shown to be fitted by the models of SNe II rather
than those of PISNe, indicating
that \sdss\ is forming stars whose mass is
$\la 30~M_\odot$. This mass range is consistent with
some metallicity studies of high-$z$ quasars
(Venkatesan, Schneider, \& Ferrara 2003).

We finally stress that
the absorption properties of dust are also important
for the observation of atoms and molecules, whose
detectability is affected by the dust extinction
(e.g.\ Shibai et al.\ 2001). Therefore, quantifying
high-$z$ extinction is crucial to discuss the
exploration of high-$z$ universe by
atomic or molecular lines.

\section{CONCLUSION}\label{sec:sum}

We have theoretically investigated the extinction curves
of young galaxies in which dust is supplied predominantly
from Type II supernovae (SNe II) and/or pair instability
supernovae (PISNe). We have adopted Nozawa et al.\ (2003)
for compositions and size distribution of grains formed in
SNe II
and PISNe. We have found that the extinction curve is quite
sensitive to the internal mixing of SNe. The extinction
curves of mixed SNe II and PISNe are dominated by SiO$_2$
and are charachterised by the steep rise from NIR to
UV because the main contribution comes from relatively
small ($\sim 0.01~\mu$m) grains. On the contrary, the
extinction curves of dust produced in unmixed SNe II and
PISNe are much flatter, because of a large contribution
from large-sized $\sim 0.1~\mu$m Si grains.

We have also derived the dust cross section per unit dust
mass. This quantity is useful to estimate the dust column
density from extinction. The UV extinction trace the
dust column density better than the optical extinction.
The extinction also affects the observability
of other molecular or atomic lines. The result of this
paper can be used to estimate the extinction effect in
high-$z$ galaxies.

Finally, our results are compared with a high-$z$
extinction curve observationally derived for \sdss\
at $z=6.2$ by Maiolino et al.\ (2004b). The
comparison favours SNe II without internal mixing as
sources of dust grains. The combination of various
progenitor mass ranging from
$\sim 10~M_\odot$ to $\sim 30~M_\odot$ explains well
the observed extinction curve. Our theoretical
extinction curves could be further utilised when
a sample of high-$z$ extinction curves is taken
by future observations.

\section*{Acknowledgments}
We thank R. Maiolino, the referee, for useful comments, and
R. Maiolino, S. Bianchi, R. Schneider, and A. Ferrara for
kindly providing us with their data on the extinction curve
of \sdss.
HH, TTI, and TTT are supported by the Japan Society for the
Promotion of Science.
TK is supported by a Grant--in--Aid for Scientific
Research from JSPS (16340051).
We fully utilized the
NASA's Astrophysics Data System Abstract Service (ADS).

\end{document}